%% file: arxiv-main.tex
\documentclass[amsmath,amssymb,superscriptaddress,
 aps,preprint]{revtex4}
\usepackage{fancyhea}
\usepackage{bm}       
\usepackage{graphicx}
\usepackage{multirow}
\usepackage{lineno}
\usepackage{tcolorbox}
\usepackage{threeparttable}
\usepackage[normalem]{ulem}
\usepackage{color}
\usepackage[colorlinks = true,
            linkcolor = blue,
            urlcolor  = blue,
            citecolor = blue,
            anchorcolor = blue]{hyperref}

\usepackage{enumitem}
\usepackage{setspace}

\begin{document}
\setstretch{1.2}


\newcommand{\conveners}{CompF4 Co-Conveners. Emails: {\tt wbhimji@lbl.gov}, {\tt rwg@uchicago.edu}, {\tt mlin@bnl.gov}, {\tt fkw@ucsd.edu}}
\newcommand{\leads}{Topic Leads. See Appendix B for details.}

\title{ \small Snowmass 2021 Computational Frontier CompF4 Topical Group Report \\\LARGE  Storage and Processing Resource Access}

\author{W.~Bhimji}\thanks{\conveners}
\affiliation{Lawrence Berkeley National Laboratory}
\author{D.~Carder}
\affiliation{Energy Sciences Network}
\author{E.~Dart}\thanks{\leads}
\affiliation{Energy Sciences Network}
\author{J.~Duarte} \thanks{\leads} 
\affiliation{University of California San Diego}
\author{I.~Fisk} \thanks{\leads} 
\affiliation{Flatiron Institute}
\author{R.~Gardner}\thanks{\conveners}
\affiliation{University of Chicago}
\author{C.~Guok}\thanks{\leads}   
\affiliation{Energy Sciences Network} 
\author{B.~Jayatilaka}\thanks{\leads} 
\affiliation{Fermi National Accelerator Laboratory}
\author{T.~Lehman}
\affiliation{Energy Sciences Network}
\author{M.~Lin}\thanks{\conveners}
\affiliation{Brookhaven National Laboratory}
\author{C.~Maltzahn}\thanks{\leads}   
\affiliation{University of California Santa Cruz}
\author{S.~McKee}\thanks{\leads}   
\affiliation{University of Michigan}
\author{M.S.~Neubauer}\thanks{\leads}   
\affiliation{University of Illinois Urbana-Champaign}
\author{O.~Rind}\thanks{\leads}   
\affiliation{Brookhaven National Laboratory}
\author{O.~Shadura}\thanks{\leads}   
\affiliation{CERN}
\author{N.V.~Tran}\thanks{\leads}   
\affiliation{Fermi National Accelerator Laboratory}
\author{P.~van Gemmeren}\thanks{\leads}   
\affiliation{Argonne National Laboratory}
\author{G.~Watts} \thanks{\leads}   
\affiliation{University of Washington}
\author{B.~A.~Weaver}\thanks{\leads}   
\affiliation{NOIRLab}
\author{F.~W\"urthwein} \thanks{\conveners}
\affiliation{University of California San Diego}

\date{\today}


\maketitle 

\tableofcontents
\clearpage 

\parindent=0pt
\parskip=10pt
\setlength{\evensidemargin}{0pt}
\setlength{\oddsidemargin}{0pt}
\setlength{\marginparsep}{0.0in}
\setlength{\marginparwidth}{0.0in}
\marginparpush=0pt

\pagestyle{fancyplain}
\addtolength{\headwidth}{\marginparsep}
\addtolength{\headwidth}{\marginparwidth}

\lhead[\fancyplain{}{\bf\thepage}]%
      {\fancyplain{}{}}
\rhead[\fancyplain{}{}]%
      {\fancyplain{}{\bf\thepage}}
\cfoot{}
\lfoot[\fancyplain{}
      {\footnotesize \sc Community Planning Exercise:  Snowmass 2021}]%
      {\fancyplain{}{}}
\rfoot[\fancyplain{}{}]%
      {\fancyplain{}{\footnotesize \sc Community Planning Exercise: Snowmass 2021}}



\input Computation/CompF04/Storage.tex

\end{document}

%% file: Computation/CompF04/Storage.tex
\section{Introduction}
\subsection{Snowmass 2021 CompF4 Scope} 
The Snowmass 2021 CompF4 topical group's scope is facilities R\&D, where we consider ``facilities" as the hardware and software infrastructure inside the data centers plus the networking between data centers, irrespective of who owns them, and what policies are applied for using them. In other words, it includes commercial clouds, federally funded High Performance Computing (HPC) systems for all of science, and systems funded explicitly for a given experimental or theoretical program. However, we explicitly consider any data centers that are integrated into data acquisition systems or trigger of the experiments out of scope here. Those systems tend to have requirements that are quite distinct from the data center functionality required for ``offline" processing and storage.

As well as submitted whitepapers, this report is the result of community discussions, including sessions in the Computational Frontier workshop~\cite{Aug2020} on August 10 -- 11, 2020, and the CompF4 Topical Group workshop~\cite{April2022} on April 7 -- 8, 2022. These workshops drew attendees from all areas of High Energy Physics (HEP), with representatives from large and small experiments, computing facilities, theoretical communities and industry.  Registered workshop participants are listed in Appendix~\ref{app:participants}. 

The community discussions quickly converged on six distinct sub-topics within this topical working group. Those include the obvious ``Storage" and ``Processing" that are already in the name of our topical group, but also potentially less obvious like ``Edge Services", ``AI Hardware", ``Analysis Facilities", and of course ``Networking". The leads for these topics are listed in Appendix~\ref{app:topicleads}. Each of these sub-topics defines itself below in its respective sections, and arrives at conclusions within its respective scope. We find that in many cases, multiple sub-areas arrive at related, or mutually reinforcing recommendations for needed action. We thus bring these together into a coherent picture, rather than just summarizing each sub-topic separately. 

\subsection{Findings and Recommendations}

The one characteristic that remains unchanged is the nature of HEP as a ``team sport'' with teams that are global in nature. These global teams will continue to require global federation of ``in-kind'' resources because each funding agency involved will make its own decisions on how to provide the required resources for a given program. The movement of data across the global research and education networks, and in/out of processing and storage facilities is thus the one characteristic that is unlikely to change.

With the slowdown of Moore's Law we expect a diversification of computing devices, architectures, and computing paradigms. R\&D is required for the community to understand how to exploit a much more heterogeneous computing and storage landscape at the facilities to contain overall costs given this slowdown. 

HEP will need to make more efficient use of facilities that are diverse both in the type of facility (e.g., dedicated grid resources; HPC and cloud) and the type of compute they have available (CPU, GPU, special purpose AI accelerators, computational storage, etc.). 

Our report calls out several areas where there are considerable opportunities to achieve these needed improvements. Significant R\&D is required to make efficient use of the diverse resources expected to be available at grid, cloud and HPC facilities, which we summarize below.
\begin{enumerate}[leftmargin=*]

\item  \emph{\bf Efficiently exploit specialized compute architectures and systems}. To achieve this will require the allocation of dedicated facilities to specific processing steps in the HEP workflows, in particular for ``analysis facilities'' (Sections \ref{sec:process} and \ref{sec:af}); designing effective benchmarks to exploit AI hardware (Section \ref{sec:aihw}); improved network visibility and interaction (Section \ref{sec:net}); and enhancements to I/O libraries such as lossy compression and custom delivery of data (Section \ref{sec:storage}). 

\item \emph{\bf Invest in portable and reproducible software and computing solutions to allow exploitation of diverse facilities.} The need for portable software libraries, abstractions and programming models is recognized across all the topics discussed here, and is especially called out in Processing (Section~\ref{sec:process}), AI Hardware (Section~\ref{sec:aihw}) and Storage (Section~\ref{sec:storage}). Software frameworks to enable  reproducible HEP workflows are also greatly needed (Sections~\ref{sec:af} and \ref{sec:edge}).

 \item  \emph{\bf Embrace disaggregation of systems and facilities}. The HEP community will need to embrace heterogeneous resources on different nodes, systems and facilities and effectively balance these accelerated resources to match workflows. To do so will require  software abstraction to integrate accelerators, such as those for AI (Section \ref{sec:aihw}); orchestration of network resources  (\ref{sec:net}); exploiting computational storage (Section \ref{sec:storage}); as well as exploiting system rack-level dis-aggregation technology if adopted at computing centers. 

 \item  \emph{\bf Extend common interfaces to diverse facilities}. In order to scalably exploit resources wherever they are available, HEP must continue to encourage edge-service platforms on dedicated facilities as well as Cloud and HPC (Section \ref{sec:edge}),  develop portable edge-services that are re-usable by other HEP projects, and exploit commonality within HEP and other sciences (Section \ref{sec:edge}). These interfaces will also need to extend into all aspects of HEP workflows, including data management and optimizing data movement (Sections \ref{sec:net},  \ref{sec:process} and \ref{sec:storage}), as well as the deployment of compute resources for analysis facilities (Section \ref{sec:af}). 

\end{enumerate}

We suggest that the funding agencies use the above recommendations in future solicitations targeting collaborative work between domain and computer science and engineering. In addition, we encourage the HEP community to be creative in using existing solicitations to write proposals that cover these areas. We note that the HEP community has been very successful in competing across all of science in these kind of solicitations, especially in NSF-CISE and NSF-OAC. On the DOE side, we encourage, for example, the community to work together with the DOE HEP office towards SciDAC proposals that cover the above recommendations. Looking into the future, we want to highlight the work by the ``National Artificial Intelligence Research Resource (NAIRR) Task Force''~\cite{ai-taskforce}. This task force is expected to conclude with its final recommendation by December 2022, and may recommend addressing some of the gaps we have identified here as it pertains to future investments in AI computational and data resources. Likewise, within the DOE there are on-going activities for post-Exascale programs around AI and ``Integrated Research Infrastructure''.  We encourage the HEP community to pay close attention to these activities and reports as they emerge.   

The rest of the report covers the detailed discussions of challenges and research directions for each topic that help derive the above recommendations. Section~\ref{sec:process} discusses processing in general. Section~\ref{sec:aihw} focuses on R\&D needed for specialized AI hardware. Storage and I/O software are discussed in Section~\ref{sec:storage}. Section~\ref{sec:af} covers research needs for analysis facilities. Edge services are discussed in Section~\ref{sec:edge}. Finally, networking challenges and research directions are presented in Section~\ref{sec:net}. 

\input{Computation/CompF04/topic_processing}

\input{Computation/CompF04/topic_AI_hardware}
\input{Computation/CompF04/topic_storage}
\input{Computation/CompF04/topic_analysis_facilities}
\input{Computation/CompF04/topic_edge_services} 
\input{Computation/CompF04/topic_networking}
\input{Computation/CompF04/Appendix}

\bibliographystyle{JHEP}
\bibliography{Computation/CompF04/myreferences}

%% file: Computation/CompF04/topic_processing.tex
\section{Processing}
\label{sec:process}

\subsection{Introduction}
Processing is the step that transforms raw data, simulation configurations or theoretical models into objects useful for analysis and discovery, and plays a central role in HEP computing.   Processing takes place in a variety of environments under different constraints.   The environments span from low latency experiment online systems, through globally distributed dedicated processing sites, to HPC and cloud allocations and opportunistic resources. 

The largest of the next generation of physics projects represent exascale science endeavours with annual data rates of exabytes to process, store, and analyze~\cite{LHCComp:2022}. Large-scale theory-based numerical simulations~\cite{osti_1581234, Kahn:2022kae, lqcd-boyle} are reaching for higher and higher precision to more accurately describe nature and uncover new science, and have been the power users of the HPC resources. Small science projects may also have big processing needs~\cite{SmallExp:2022wp}. To meet these challenges we will need to exploit a changing landscape of new hardware and new techniques.  More than 20 years ago science switched to the x86 processor and commodity computing.   Today there are accelerated processor architectures like GPU, FPGAs, and TPUs, which show dramatic performance improvements for certain types of calculations. Low-power general purpose ARM processors are appearing in devices from iPads to supercomputers. Exploiting new architectures requires investment in software design and portability but opens access to new resources like HPC facilities.  New techniques like AI/ML, advanced data analytics, and digital twins~\cite{digital_twins} change how we think about science processing and simulation as well as the computing and IO requirements.

\subsection{Challenges} 
The evolution of the computing landscape introduces many challenges. The increasing prevalence of heterogeneous computing systems makes it essential to adapt the existing software stacks that have been largely developed for homogeneous CPU-based systems~\cite{FASER:2022yqp, Kahn:2022kae}.  The growth of HPC and Cloud computing systems concentrates unprecedented computing resources away from the scientific instruments and the custodial storage, which places new demands on data access and networking~\cite{girone_maria_2020_3647548}. The introduction of new techniques like AI and ML can change the performance of the workflow, but can also change the resource balance with significant processing and data access needed for training before the workflow processes data.  The IO requirements for the inference step can be much larger than traditional workflows and there is often specialized hardware as described in Section~\ref{sec:aihw}. 

The technical processing challenges are listed below.

\begin{itemize}
   
    \item {\bf Heterogeneous Hardware:} Scientific code is the result of contributions from many people of varying skills over many years. Even maintaining and optimizing for a single platform has been challenging. The increased diversity of accelerated hardware architectures that are deployed for processing is exacerbating this challenge. The GPU market now has three players: NVIDIA, AMD and Intel, each of which has its own native programming API: CUDA for NVIDIA, HIP for AMD and SYCL/OneAPI for Intel. It is impractical to rewrite the vast HEP  software stacks for each platform. A sustainable solution that incorporates software portability, productivity and performance is critically needed to exploit the heterogeneous computing resources that will be widely available in the next decade. 
    \item {\bf Resource Interfaces:} The WLCG (Worldwide LHC Computing Grid) and the OSG (Open Science Grid) have served the data intensive science community for more than a decade. The protocols and interfaces to connect to grid sites have functioned and scaled, but the integration of new resources like HPC and clouds sites is a new technical challenge. The HPC facilities have stricter cyber-security requirements and Authentication and Authorization Infrastructure (AAI) needs. 
    \item {\bf Resource Description:} The increase in the use of heterogeneous architectures and the integration of HPC and cloud resources dramatically increases the diversity of information needed to describe resources and make intelligent scheduling decisions.
    \item {\bf Provisioning and Policy:}  Increasing the use of HPC and clouds opens new resources but introduces new challenges for how they are  provisioned and consumed. HPC facilities typically make awards for fixed allocations during a period of time. The time scale for a computing award might range from months to a year, but is significantly different from the relationships established between the dedicated grid sites, which might last for decades. Clouds add the additional complexity of having a cost per use.   Both HPC and Clouds are fixed resources, either due to allocation or budget, and this places challenges on how to predict usage and enforce experiment priorities. 
    \item {\bf Data Management and Delivery:} Scientific computing has traditionally maintained a reasonably strict coupling between processing and storage resources.   Data is moved to dedicated storage and accessed locally with only a minority share, if any, of the data streamed. The addition of non-dedicate processing resources like HPC and clouds places challenges on the data management system to be more dynamic.  The proposals to use \textit{DataLake} style data management models places demands on the networking, data federation and data caching infrastructure. 
    
    Additionally, the increased use of accelerated hardware solutions can improve the performance of processing, but it also increases the challenges of data management.  If application performance is increased by a factor of ten, the I/O must scale commensurately. The interfaces to storage need to be evaluated in the presence of accelerated architectures and workflows. 
    \item {\bf Impact of ML-based Processing:} The adoption of Machine Learning-based workflows in processing-intensive applications has the potential to dramatically improve the application throughput, but introduces challenges in the balance of resources and the types of computing needed. ML training is processing intensive and needs to be performed before the real data workflows can be performed. It is potentially a good application for HPC sites.  The inference step requires much less computing, but can benefit from dedicated hardware like FPGAs.

\end{itemize}

\subsection{Research Directions}
In order to overcome the challenges facing processing for scientific computing in the next decade, we need to establish research directions and make investments.  We should take the opportunity to rethink our historical choices and evaluate what are the right decisions to best complete our work given the changing technology landscape. 

\subsubsection*{What are the best processing facilities for HEP research in the future?}
The first research question that needs to be answered is what are we optimizing for when we design our processing systems and decide what resources to use.  We should take the opportunity of the planning exercise to assess the efficiency of ways of working without the constraint of what is currently deployed. Possible metrics for establishing the ``best" solution include the following:
\begin{itemize}
    \itemsep-0.3em

\item Overall cost
\item Utilization of existing infrastructure
\item Time to results 
\item Familiarity and comfort level of the user community
\item Carbon footprint
\item Minimized effort
\item Synergies with other science activities or industry
\end{itemize}

The most important aspect is to decide in advance what are the criteria that will be used in making choices. The landscape is changing and there are many new elements since many HEP computing models, such as LHC, were designed decades ago. The addition of clouds, HPC, and heterogeneous architectures open many opportunities but all come with benefits and costs.  It is unlikely that one solution or optimization will apply to the entire research program over a decade, but the process to establish what is important and to justify what choices were made is common.

\subsubsection*{Research Areas}
In addition to the big question of what we are optimizing for, there are a number of more specific research directions needed that will serve as input to the optimization question and help the field navigate the changing landscape.

\begin{itemize}
    \item {\bf Use of heterogeneous architectures:} The use of heterogeneous hardware architectures including accelerated co-processors has traditionally involved specialized skills and a redesign of the application to achieve reasonable performance. Recently unified programming models and portability libraries are opening the possibility of a single code base that runs with reasonable efficiency on multiple architectures. Adding new architectures can be done once in the portability layer, improving code maintainability.  Additional research and a systematic approach to move the field to be more flexible in terms of supported hardware platforms is needed~\cite{Bhattacharya:2022qgj,Jones:2022ycw}. 
    \item {\bf Evolution of resource sharing and provision:}  The ability to integrate new hardware architectures and to deliver data to non-dedicated resources will enable growth in the resource pool with the addition of clouds, HPC sites, and other opportunistic facilities.   The typical resource provisioning of annual pledges with the expectation of a commitment over the life of an experiment will not necessarily work for these new classes of resources.   Research is needed in how we might burst to much larger resources enabling provisioning for peak and execution of fixed duration computing activities.  Evaluations are needed into what percentage of processing activities could map efficiently onto HPC and cloud allocations. Negotiations are also needed with the HPC and Cloud providers if alternative longer term provisioning would be acceptable in some cases. 
    \item {\bf Evolution of data access:} In data intensive science it is impossible to separate processing and data access.  Traditionally, this has meant coupling data storage and processing infrastructures.   As we evolve to exploit large-scale HPC and cloud computing resources we need to explore data access solutions that are more dynamic and make efficient use of caching and the network.  The \textit{DataLake} models proposed for the LHC are moving in the right direction but need to be able to scale to deliver 10s of Petabytes daily to remote processing sites and potentially to export similar data volumes. 
    \item {\bf Evolution of interfaces:} One of the big successes of the grid was a common set of interfaces for processing, storage access, and information services. Those services relied on very similar destination hardware and provisioning and accounting expectations.  With a more diverse landscape including non-dedicated sites, large HPC allocations, heterogeneous hardware solutions, and rented computing services, we need to develop an enhanced set of interfaces that scale both in size and environment complexity.   
    \item {\bf Modifying computing models:} LHCb and ALICE, even for LHC Run3, have moved to a largely triggerless configuration where most the offline processing is performed in nearly real-time. The smaller reconstructed objects are stored.  This pushes many traditional offline workflows into the online environment including some analysis steps. This technique has efficiency gains in processing because the online and offline elements are not duplicated and can save significantly in storage if only synthesized data formats are retained~\cite{Bartoldus:2022zlc}. It increases the risks in offline processing because there are not necessary resources or raw data formats available to recover from a problem in the data reconstruction. A general assessment of the benefits and risks of moving more workflows to real-time, single pass execution should be performed. 
\end{itemize}

\subsubsection*{Types of Computing Resources}
Building the original grid infrastructure was a large multi-national investment over years, but the resulting infrastructure has enabled the distribution of computing sites to facilitate the efficient use of local computing investments and has provided the LHC experiments with processing capabilities from day 1.  It has also demonstrated the ability to move and process data globally and the need to treat processing, storage, and networking as equal partners in sustaining a computing model. To build the next generation of processing infrastructure, one that allows a rich diversity of hardware architectures and includes contributions from HPC sites, institutional clusters and clouds, will be a significant investment also.   The project can and should be divided by technical area: application software, services and interfaces, data management, etc. Forming projects that include a mix of large and small experiments with different requirements and workflows will help find common technical solutions.  

\subsection{Recommendations}
We conclude this section with the following recommendations to  meet the increasing HEP processing needs in the next 10-15 years. 
\begin{itemize}

    \item HPC facilities should revisit their resource access policies to allow more flexible allocations and job executions. This, coupled with new authentication and authorization models, will allow more HEP projects to benefit from the large computing facilities. 
    \item Investment in software development effort is key to maximize the efficient utilization of diverse processing resources. In particular, research and development of portable software solutions is critical for a sustainable software ecosystem in light of the evolving and increasingly diverse hardware architectures. 
    \item Research is needed to determine the tradeoff between dedicated HEP computing facilities and general-access computing facilities such as the HPC center, Grid and Cloud resources. 
    \item Infrastructure development will be needed to support better data management frameworks across different types of facilities.  
\end{itemize}

%% file: Computation/CompF04/topic_AI_hardware.tex
\section{AI Hardware}
\label{sec:aihw}

\subsection{Executive Summary}

Artificial intelligence and machine learning are becoming increasingly prevalent in all stages of data processing, generation, and simulation across HEP to gain deeper insight into data and accelerate discovery.  
With uniquely massive data sets in science and high data acquisition rates, high-performance and high-throughput computing resources are an essential element of the experimental particle physics program. These experiments are constantly increasing in both sophistication of detector technology and intensity of particle beams. 
With growing data rates and total volumes and rapidly developing AI techniques pushing the computing capacity of HEP, more efficient hardware architectures specially designed for AI computations are a clear path to mitigating these effects.  
Similarly, theory calculations and physics simulations are also growing in complexity and powerful AI algorithms benefit from powerful AI hardware.  

In this section, we focus on the application of novel AI hardware for accelerating offline data processing.  However, there are closely related focus areas with similar themes such as machine learning (CompF03), instrumentation trigger and data acquisition (IF04), and electronics/ASICS (IF07).  

\begin{figure}[tbh!]
    \centering
    \includegraphics[width=1.0\textwidth]{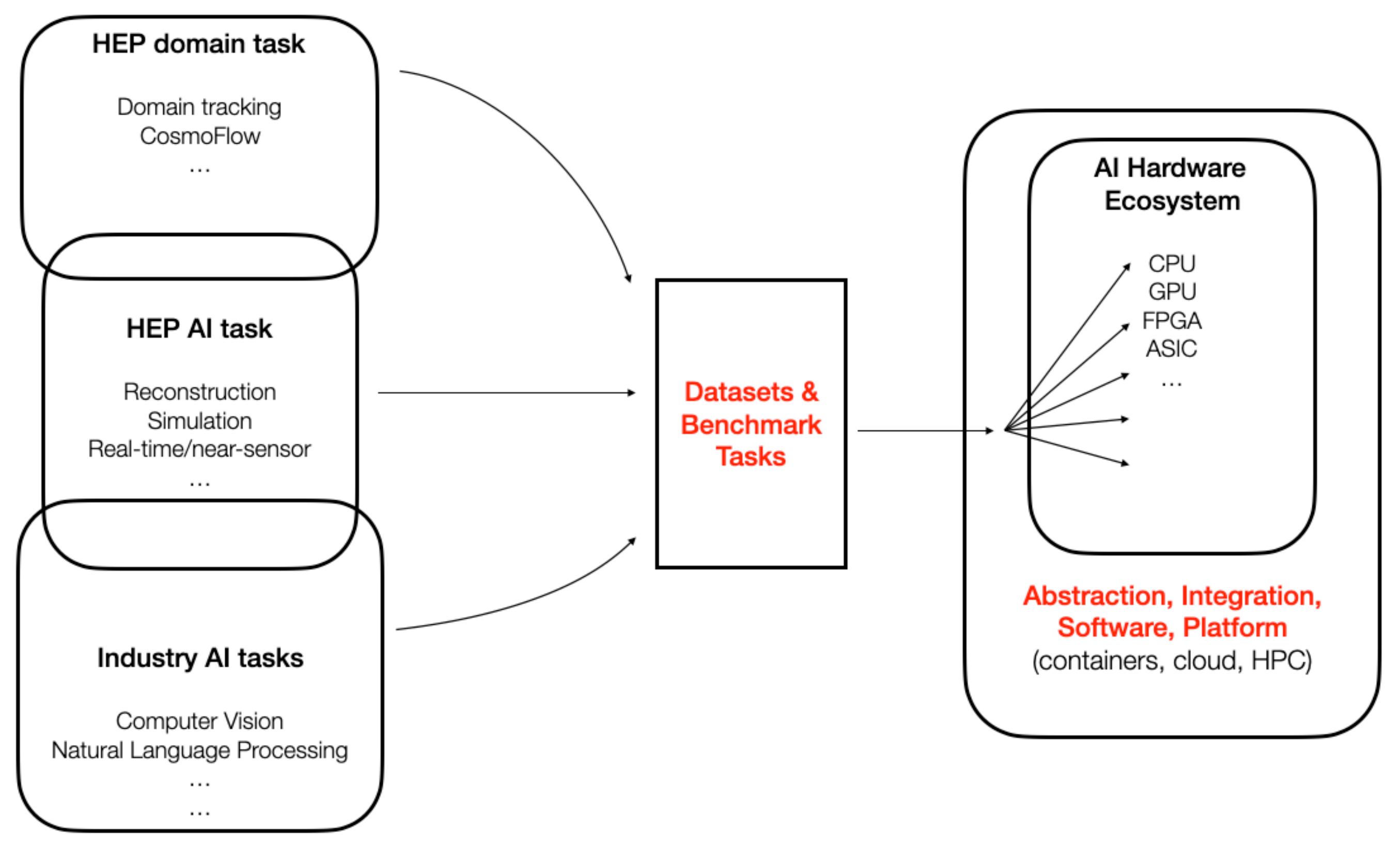}
    \caption{AI hardware ecosystem including scientific and industry tasks---we highlight the connection of hardware with HEP challenges including areas for development, benchmarking and abstraction }
    \label{fig:sum}
\end{figure}

AI hardware has been developing rapidly with many technologies recently becoming available and others anticipated.  
This space includes traditional CPUs, as well as GPUs which are the current standard for AI workloads, but a number of emerging hardware platforms such as FPGAs, ASICs, and deep learning processors (DLPs) which are specialized architectures for AI and includes both traditional CMOS and beyond-CMOS technologies.  
While HEP is not a driver for the advancement of these numerous technologies, it is important to systematically study the landscape of AI hardware and understand which architectures are best suited for various important and unique HEP AI tasks and thus its science drivers.  
To that end, we have identified two main areas of development, shown in Fig.~\ref{fig:sum} that would inform and improve the adoption of AI hardware in HEP computing workloads.  
First, it is important to establish curated datasets and AI benchmark tasks with robust metrics on which different AI hardware can be evaluated.  
These AI benchmark tasks should highlight HEP complementary to standard industry benchmarks. Second, because AI hardware is continually advancing, there may not ultimately be a single solution and it is important to develop software and computing infrastructure to efficiently integrate and abstract, e.g. ``as a service'' access, any number of AI hardware platforms into HEP computing workflows.

\subsection{HEP Computing Challenges}

As HEP computing ecosystems grow in scale and complexity, new data processing and reduction paradigms need to be integrated into the computing infrastructure design. 
Fortunately, this coincides with the rise of machine learning (ML), or the use of algorithms that can learn directly from data. 
Recent advancements demonstrate that ML architectures based on structured deep neural networks are versatile and capable of solving a broad range of complex scientific problems. 
While each scientific application is unique, there are large overlaps in data representations and computing paradigms.  
In Ref.~\cite{10.3389/fdata.2022.787421}, a summary of scientific needs and science drivers are presented for a number of HEP applications such as DUNE, the LHC experiments, cosmology surveys, intensity frontier experiments, and accelerator operations.  Furthermore, there is also discussion of other non-HEP applications and how they dovetail with HEP workloads.  

\begin{figure}[tbh!]
    \centering
    \includegraphics[width=0.8\textwidth]{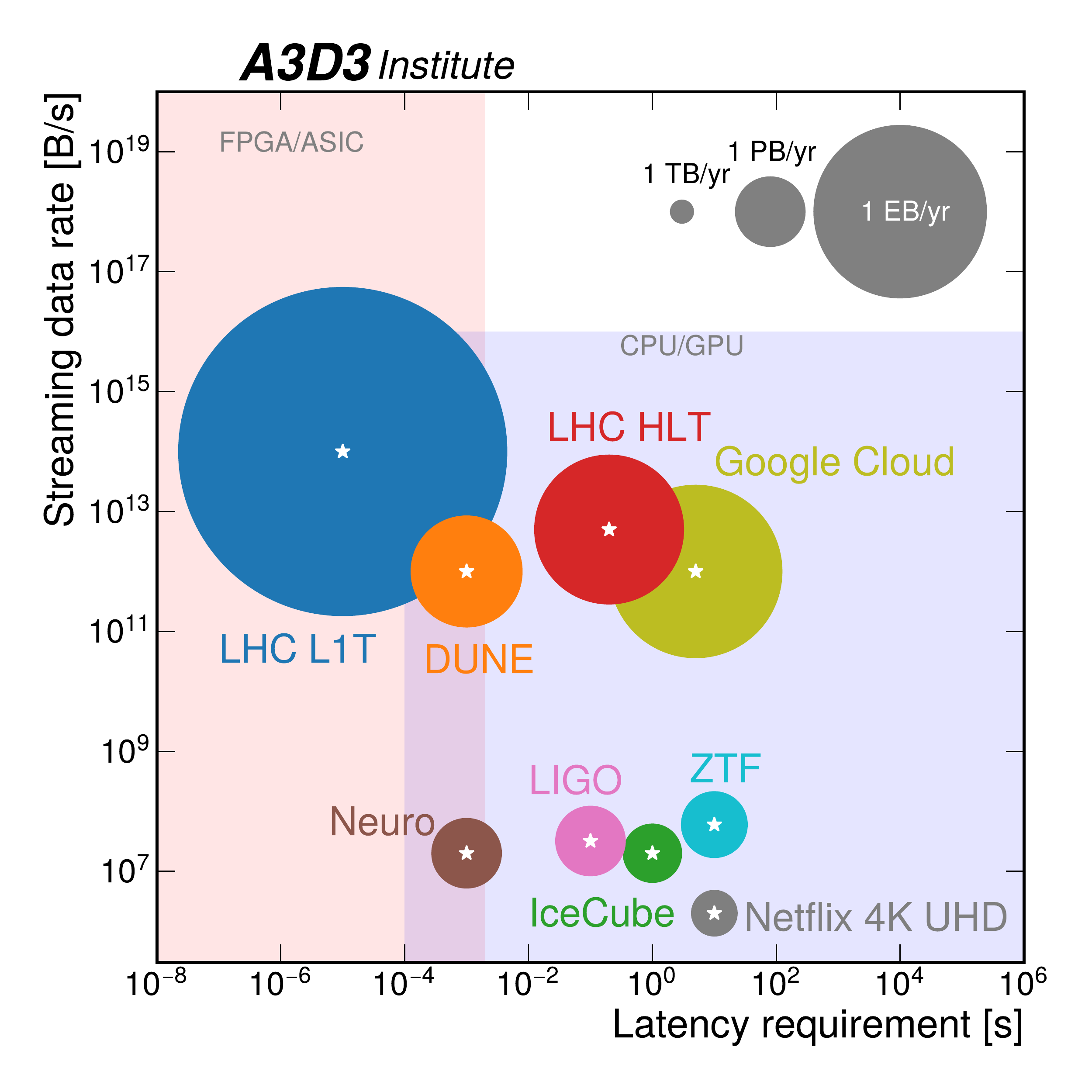}
    \caption{Plot of the streaming data rate in bytes per second and latency requirements in seconds for various experiments. Points of comparison from industry and other scientific fields are also included. 
    The size of the bubbles represents the total per year data volume. Taken from Ref.~\cite{a3d3}.}
    \label{fig:a3d3}
\end{figure}

The unique aspect of the HEP computing challenges that go beyond traditional industry workloads is their combination of data rates, latency/throughput requirements, data volumes, and data representations.  The first three can be neatly summarized in Fig.~\ref{fig:a3d3} which shows how HEP computing workloads compare to industry applications and demonstrate how the requirements are similar or can even exceed those of traditional benchmarks.  

Beyond the computing system requirements, HEP workloads also can have a variety of unique AI data representations and computing motifs.  
In Ref.~\cite{10.3389/fdata.2022.787421}, for example, there is a discussion of overlapping and common data representations across a variety of domains including expert domain features as inputs, spatial data in regular (Cartesian) space or sparse, irregular point clouds, temporal data, and spatiotemporal data.  Furthermore, the input size and their batches are important as well, including if batches are ragged (varied) for each inference or training graph, the total size of the model, and the output structure and size.  

\subsection{Hardware Taxonomy}
Just as no single ML architecture is the most appropriate for all problems, no single hardware architecture will be optimal for addressing every physics use-case effectively.
As the technology and the field evolve, so too will the methods most optimized to different use-cases.
Hardware that can be faster and more efficient than traditional CPUs for inference is one possibility for reducing the overall computing load of ML.

We present a high-level taxonomy of these hardware architectures and discuss their relevant characteristics when it comes to the acceleration of machine learning workloads. 
This is essential to understand how they will differ in their execution behavior, what it takes to leverage their unique features and how they can potentially benefit from previously introduced optimization techniques. 

\begin{figure*}[tbh!]
\centering
\includegraphics[width=0.9\linewidth]{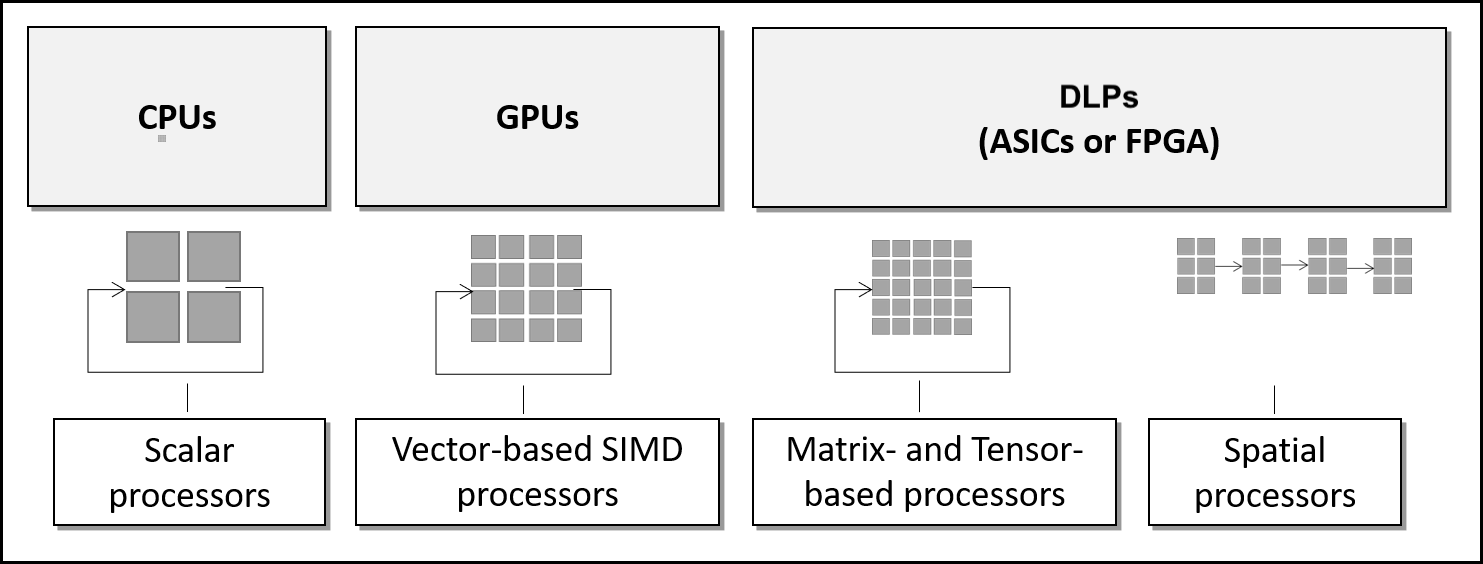}
\caption{Taxonomy of compute architectures, differentiating CPUs, GPUs, and DLPs}
\label{fig:tax}
\end{figure*}

A broad range of hardware architectures to deploy machine learning algorithms exists today.
We can broadly characterize them by the following criteria:
\begin{itemize}
    \itemsep-0.3em
    \item Basic type of compute operation
    \item Inherent support for specific numerical representations
    \item External memory capacity (which is mostly relevant for training workloads)
    \item External memory access bandwidth
    \item Power consumption in the form of thermal design power (TDP)
    \item Level of parallelism in the architecture and the degree of specialization
\end{itemize}
 
As is shown in Figure~\ref{fig:tax}, we classify the compute architectures into scalar processors (\textbf{CPUs}), vector-based processors (\textbf{GPUs}), and so-called deep learning processors (\textbf{DLPs}), although realistically these categories blend to some degree. 
DLPs are specialized for this application domain whereby we distinguish the more generic matrix- or tensor-based processor and a spatial processing approach. 
DLPs can be implemented with either ASICs or FPGAs. 
All of these architectures will be discussed individually below.

\begin{itemize}
    \itemsep-0.3em
    \item CPU: CPUs are widely used for ML applications and are viewed as largely serial or scalar compute engines (high-end variants may have up to 10s of cores). 
    They are optimized for single-thread performance, with implicitly managed memory hierarchies (with multiple levels of caches), and support floating point operations (FP64 and FP32) as well as 8bit and 16bit integer formats with dedicated vector units in most recent variants.
    \item GPU: GPUs are SIMD-based (Single Instruction, Multiple Data) vector processors that support smaller floating point formats (FP16) natively, as well as fixed point 8-bit and 4-bit integer formats more recently, and have a mix of implicitly and explicitly managed memory. NVIDIA GPUs are some of the most popular hardware targets for machine learning---others include AMD and Intel GPUs.  
    \item FPGA/ASIC: FPGA and ASIC customize hardware architectures to the specifics of a given application. 
    Fig.~\ref{fig:tax} shows two such architectures (spatial dataflow or matrix of processing elements). 
    They can be adapted in all aspects to suit a use-case's specific requirements including IO capability, functionality, or even to suit specific performance or efficiency targets. 
    FPGAs can be reprogrammed whereas ASICs are fully hardened.  
    Examples can be far-ranging, e.g. Google TPU, Intel Habana Goya, Cerebras WSE, Graphcore IPU, IBM True North, Mythic Analog Matrix Processor, etc.
    \item Beyond CMOS: This includes again, a wide-range of exploratory technologies for efficient Vector-by-Matrix Multiplications including photonics, floating gates, emerging memory technologies, hyperdimensional computing, and more as well as dedicated technologies for spiking or neuromorphic neurons with metal-oxide or diffusive memristors. 
\end{itemize}

\subsection{AI Ecosystem and Integration}

\subsubsection*{AI Benchmarking}

Among existing AI benchmarks, the community-driven MLPerf benchmarks from MLCommons~\cite{mattson2020mlperf} are well-established.
The benchmarks are run under predefined conditions and evaluate the performance of training and inference for hardware, software, and services. 
MLPerf regularly conducts new tests and adds new workloads to adapt to the latest industry trends and state of the art in AI across various domains including high performance computing (HPC)~\cite{farrell2021mlperf}, datacenter~\cite{inferencedatacenter2021}, edge~\cite{inferenceedge2021}, mobile~\cite{reddi2020mlperf}, and tiny~\cite{banbury2021mlperf}. 
Additionally, BenchCouncil AIBench is a comprehensive AI benchmark suite including AI Scenario, Training, Inference, Micro, and Synthetic Benchmarks across datacenter, HPC, IoT and edge~\cite{benchcouncil}. 
Other benchmarks have also been developed by academia and industry. 
Additional examples of prior art and initiatives include AI Benchmark ~\cite{ignatov2019ai}, EEMBC MLMark~\cite{torelli2019measuring}, AIMatrix~\cite{aimatrix}, AIXPRT~\cite{aixprt}, DeepBench~\cite{deepbench}, TBD~\cite{zhu2018benchmarking}, Fathom~\cite{adolf2016fathom}, RLBench~\cite{james2020rlbench}, and DAWNBench~\cite{coleman2017dawnbench}. 

However, scientific applications (i.e., cosmology, particle physics, biology, clean energy, etc.) are innately distinct from traditional industrial applications with respect to the type and volume of data and the resulting model complexity~\cite{farrell2021mlperf}. 
The MLCommons Science Working Group~\cite{mlcommonsscience} has a suite of benchmarks that focus on such scientific workloads including application examples across several domains such as climate, materials, medicine, and earthquakes.
SciMLBench~\cite{thiyagalingam2021scientific} from the Rutherford Appleton Laboratory is another benchmark suite specifically focused on \textit{scientific machine learning} and aimed towards the ``AI for Science" domain. 
The suite currently contains three benchmarks that represent problems taken from the material and environmental sciences. 
MLPerf HPC and AIBench HPCAI500 are two more benchmarks that include scientific workloads. 
In general, HPC is being leveraged by the scientific community for accelerating scientific insights and discovery. 
MLPerf HPC aims to systematically understand how scientific  applications perform on diverse supercomputers, focusing on the time to train for three representative scientific machine learning applications with massive datasets (i.e., cosmology, extreme weather analytics, and molecular dynamics).
Similarly, AIBench HPCAI500 also includes a benchmark on extreme weather analytics.

Within HEP, there have been some initial efforts to define AI benchmarks such as the top-tagging~\cite{toptagging} and Kaggle tracking ML challenge~\cite{Amrouche:2021tio}, the latter being one of the few which that emphasizes balancing the accuracy of the solution with the speed of inference. 
We can leverage experiences with these organizations to build out more HEP-specific AI benchmarks that define metrics for both physics performance and computing efficiency.
As discussed above, \textbf{this will be an evolving and dynamic program that should be sustained with the evolution of hardware.}

\subsubsection*{Software Abstraction and Integration}

\begin{figure*}[tbh!]
\centering
\includegraphics[width=0.7\linewidth]{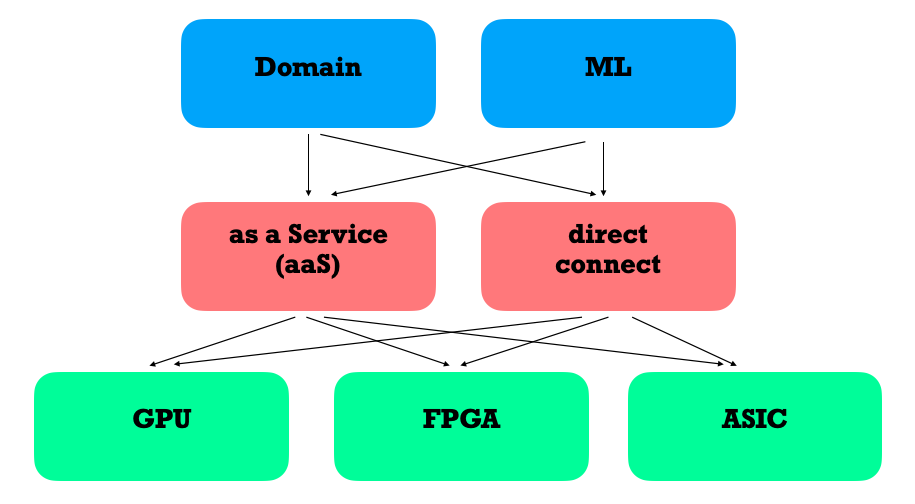}
\caption{Paths for deploying AI coprocessors for HEP algorithms}
\label{fig:paths}
\end{figure*}

Because flexibility is required in evaluating constantly evolving AI hardware for a wide array of evolving HEP tasks, there are a number of paths to deploying coprocessor hardware for HEP use-cases.  
This is illustrated in Fig.~\ref{fig:paths} where either domain or machine learning algorithms can run on any number of technologies (GPU, FPGA, ASIC, etc.).  
We classify how the coprocessor hardware is connected to the CPU host system as either ``direct connect" or ``as a service (aaS)."  
The first is more optimal, typically running bare-metal applications, while the latter is abstracted and can be more versatile in their deployment---often times not co-located with the CPU host processor.  

We present a number of considerations when designing a system to deploy coprocessor AI hardware and note that investment is required to develop technologies that can accommodate a variety of hardware.

\begin{itemize}
    \itemsep-0.3em
    \item \textbf{Flexibility}: Allowing multiple clients to connect to multiple coprocessors enables many arrangements to ensure optimal usage of all devices.
    \item \textbf{Cost-effectiveness}: Related to flexibility, using coprocessors optimally reduces the number of coprocessors that must be purchased to support algorithm inference.
    \item \textbf{Symbiosis}: Where possible, facilitate the use of existing industry tools and developments, rather than requiring HEP software developers to reimplement common tasks such as ML algorithm inference repeatedly for different ML frameworks, coprocessors, etc.
    \item \textbf{Simplicity}: Modules only implement conversions of input and output data, which reduces the amount of code necessary to develop and maintain in order to perform ML algorithm inference.
    \item \textbf{Containerization}: Model abstraction and containerization keeps the ML frameworks separate from the experiment software framework, eliminating the significant workload needed to integrate two software systems that each have their own complicated dependencies.
    \item \textbf{Portability}: Related to containerization, enable experiment software workflows to swap between CPUs, GPUs, FPGAs, and other coprocessors without any code changes including the choosing the ML framework with no other modifications.
\end{itemize}

Trade-offs for these considerations should be compared when considering direct connect versus aaS paradigms.  
As an example of on-going R\&D, Services for Optimized Network Inference on Coprocessors (SONIC) is a software design pattern to integrate a client-server approach for inference as a service into experiment software frameworks (which are usually based on C++).
It offers useful abstractions to minimize dependence on specific features of the client interface provided by a given server technology.
SONIC has been implemented in the CMS software~\cite{Duarte:2019fta,Krupa:2020bwg,Rankin:2020usv} and in LArSoft for protoDUNE~\cite{Wang:2020fjr};
it is being explored by other experiments including ATLAS.
Existing implementations of SONIC~\cite{Krupa:2020bwg,Wang:2020fjr} focus on the open-source Triton inference server from NVIDIA~\cite{Triton}. 
This enables the automatic portability that is a key advantage of the SONIC approach.
In the future, other client-server technologies such as the interprocess communication (IPC) provided by Apache Arrow~\cite{arrow} could be considered.

%% file: Computation/CompF04/topic_storage.tex
\section{Storage}
\label{sec:storage}

With ever increasing data rates of future HEP experiments such as at the HL-LHC, Input/Output and storage of both RAW and derived data will become more challenging and costly. And as processing of HEP data moves to new architectures, I/O and storage infrastructure needs to adapt to these changes.

\begin{figure}[tbh!]
  \centering
  \includegraphics[width=0.45\linewidth]{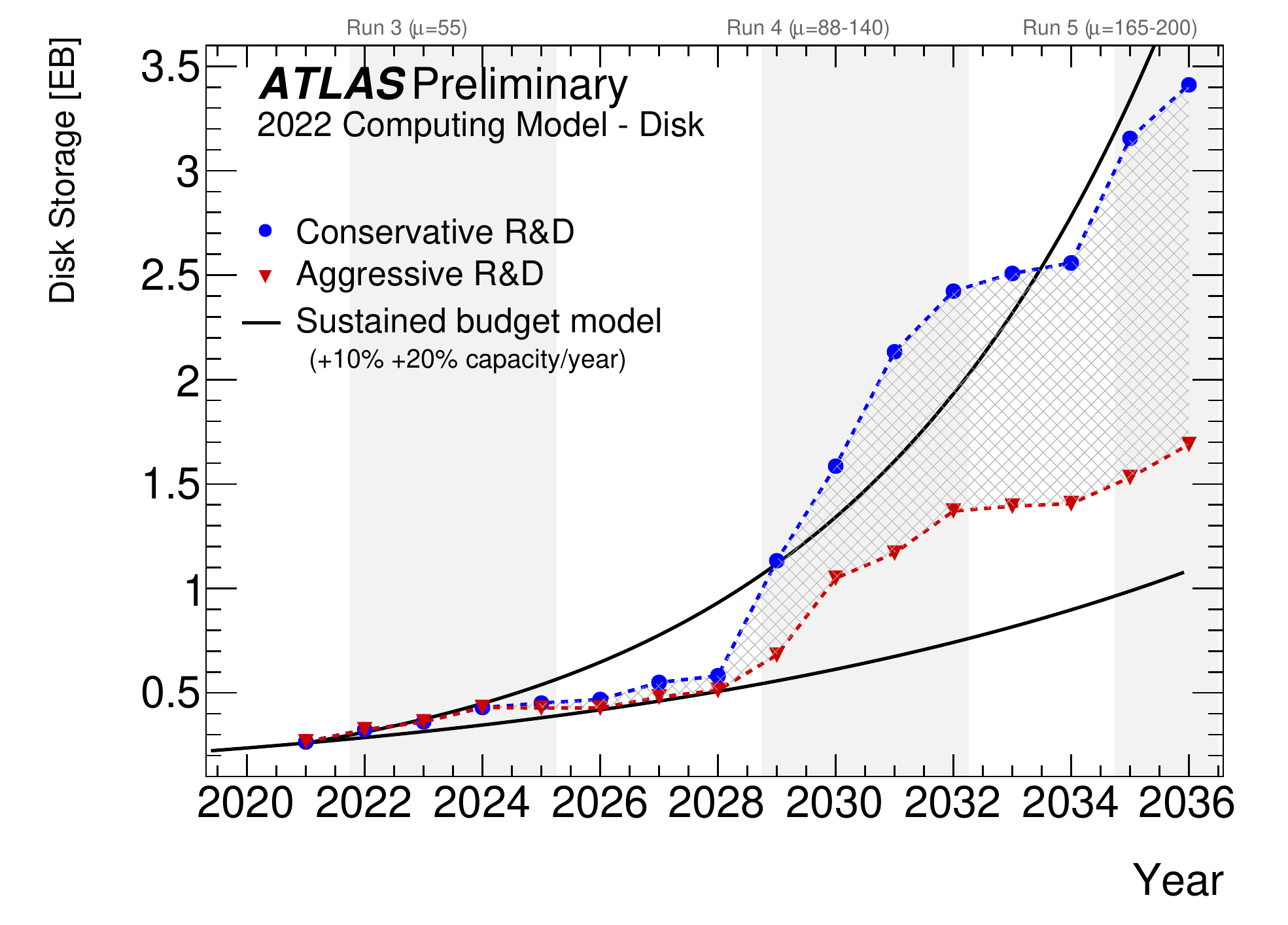}
  \includegraphics[width=0.5\linewidth]{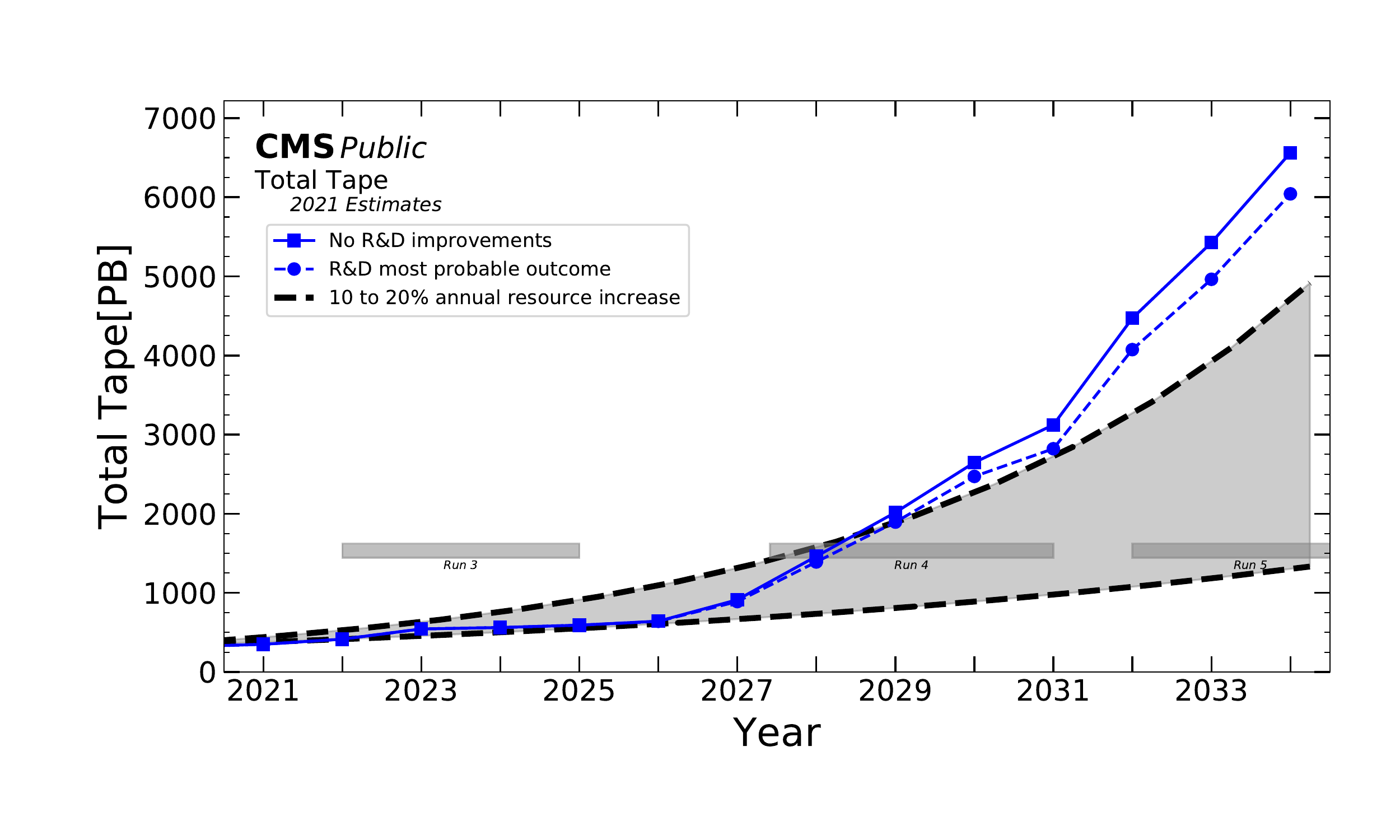}
  \caption{Left: The expected disk storage needs of ATLAS~\cite{atlasStorage} in Exabytes as a function of time showing storage available under a sustained budget scenario (black curves) and required storage for both conservative (blue dots) and aggressive (red dots) R\&D, the latter on including lossy compression of derived data. Right: The expected tape storage needs of CMS in Petabytes as a function of time assuming no R\&D improvements (solid blue) or probable R\&D improvements (dashed blue). Overlaid are expected tape resources extrapolating from 2018 pledges and assuming a 10-20\% increase in budget (shaded region).  \label{fig:storage}}
\end{figure}

\subsection{Storage Media}
Storage technologies in use by HEP can be considered in one of three categories: magnetic tape, rotating magnetic hard drives, and solid-state storage. While the cost and performance of each technology continue to evolve, the relative hierarchy in terms of cost per byte and latency continue to be the same with tape on the bottom and solid state on the top in terms of cost and the same hierarchy inverted in terms of latency. Other technologies are not foreseen to be a factor for HEP in the coming decade with optical storage not practical at the necessary scales and technologies like holographic memory and DNA storage being anticipated much further in the future. In the timescale covered by this Snowmass report, the needs and purchasing influence of hyperscaler providers will primarily govern the storage landscape. 

\subsubsection*{Tape}
Magnetic tape continues to be the backbone of archival data storage in HEP with sites such as CERN, BNL, and Fermilab managing several hundred petabytes of active tape storage each. The areal density of data on tape is far from reaching physics-determined limits. The future of tape is thus much more driven by market forces than by more fundamental limits. Large-scale tape storage deployments are challenging to plan and operate with three basic components: the storage media in the form of tape cartridges, tape drives, and robotic tape libraries. Additionally, dedicated networking and compute servers may be needed. Thus, HEP tape storage tends to be centralized at a few sites per participating country; a trend that is unlikely to reverse in the future. Tape cartridge capacity, which currently peaks at approximately 20 TB, is expected to increase much faster than tape drive bandwidth, which is currently at approximately 400 MB/s. Thus, aggregate bandwidth is likely to be a bigger factor in estimating tape costs at HEP sites in the future than overall storage capacity.

\subsubsection*{Rotating Disk}
Despite its rapid disappearance in the consumer segment, rotating hard disk drives (HDDs) continue to provide the bulk of active storage in enterprise data centers. HDDs also provide nearly all of the storage in small to medium HEP computing sites as well as the nearline storage and cache for tape-enabled archival storage sites. Unlike tape storage, HDD areal density has largely stagnated in recent years with manufacturers turning to adding platters to increase drive capacity. Perpendicular Magnetic Recording (PMR), which is used in most HDDs, is unlikely to increase drive capacity beyond 20TB. Manufacturers are turning to technologies such as Heat-Assisted Magnetic Recording (HAMR), Microwave-Assisted Magnetic Recording (MAMR), and Shingled Magnetic Recording (SMR) to go beyond the limitations of PMR. Hard drives with all of these technologies have been brought to market as of 2022. The cost per byte of HDDs, while lagging behind that of tape, continues to be less than that of solid-state storage, despite the cost of the latter dropping substantially over the past decade. 

\subsubsection*{Solid-State Storage}
Solid-state storage now dominates the consumer market for storage both with portable and desktop computers as well as exclusively for mobile devices. Despite the vast increase in production of solid-state storage driven by this demand, a significant price gap between it and HDDs persists with an approximately order of magnitude gap in terms of price per byte. Thus, much of the enterprise storage market continues to be driven by HDDs, including at the large hyperscalars. Solid-state storage usage continues to be small in HEP outside of system/local disks and specialized caches. While solid-state storage allows greater aggregate throughput as compared to HDDs, its most significant performance advantage is in IOPS. As HEP potentially moves away from monolithic architectures for computing, the use of high-speed solid-state storage will considerably benefit platforms devised for end-user analysis which are often more limited by IOPS. 

\subsection{Storage and I/O Software} 
\subsubsection*{ROOT}
ROOT has been the primary format for storage of experimental HEP data since well over two decades and today experiments store over 1 Exa Byte of data within ROOT's TTree storage type.
Over the next five years, ROOT will undergo a major I/O upgrade of the event data file format and access API and provide a new storage type: RNTuple ~\cite{rootRNTuple}, which is expected to eventually replace TTree.
The reasons for this transition are substantial performance increase expectations: 10-20\% smaller files, 3--5 times better single-core performance, because RNTuple is developed with efficient support of modern hardware (GPU, HPC, Object Stores, etc.) in mind (built for multi-threading and asynchronous I/O).
In addition, RNTuple promises: native support for HPC and cloud object stores, systematic use of check-summing and exceptions to prevent silent I/O errors and inclusion of lossy compression algorithms.

\subsubsection*{Data Storage for HEP Experiments in the Era of High-Performance Computing}
Processing for future HEP experiments, such as HL-LHC, faces large challenges due to processing cycles, and porting workflows to HPC systems is being considered as mitigation ~\cite{hpcStorage}. In addition to data offloading to compute accelerators such as GPUs, this approach requires scalable and efficient data storage and input/output. HPC systems often feature custom storage infrastructure, often with multiple layer hierarchy (such as parallel file system and burst buffer) that may be used more efficiently using more HPC native storage software such as HDF5.

The last few decades have been dominated by the grid computing ecosystem where ROOT has been used by most of the HEP experiments to store data. Initial explorations with HDF5 have begun using the ATLAS, CMS and DUNE data.

An I/O test framework has been under development by the HEP Center for Computational Excellence (HEP-CCE) that supports the study of scaling of the I/O performance with different data-formats on different systems by looking at memory usage, file size, compression and storage software like ROOT and HDF5.

Studies done by the HEP-CCE project have relied on ROOT serialization for typically complex HEP event data models to be stored as binary objects in HDF5, as HDF5 does not provide the same near-automatic C++ type support. Storing binary objects requires having their original type, size and location/offset as well. Different data mapping methods have been investigated, such as storing individual data objects in separate HDF5 Datasets or accumulating the complete event content into a single HDF5 Dataset. HEP-CCE also has developed a prototype exercising HDF5’s collective output capability, writing to the same file from multiple processes.

\begin{figure}[tbh!]
  \centering
  \includegraphics[width=\linewidth]{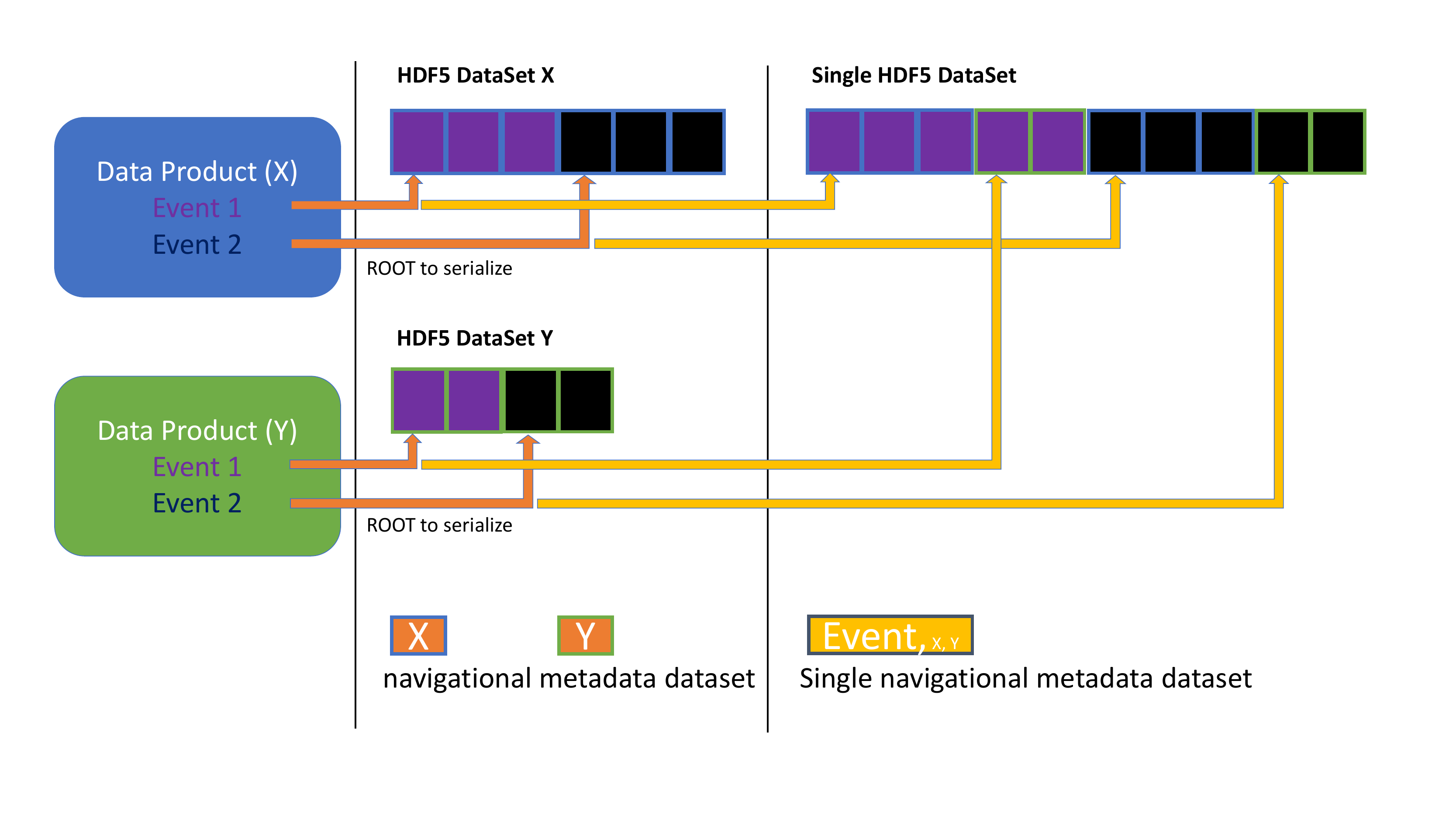}
  \caption{Two different ways of mapping HEP data to HDF5. Leftmost two boxes represent the X and Y data products of the events. The center column shows the one-to-one mapping where data products are stored in individual HDF5 datasets after ROOT serialization. The rightmost column shows the mapping where all attributes of an event are accumulated after serialization and stored in a single HDF5 dataset \label{fig:hdf_mapping}}
\end{figure}

Simpler event data models, such as raw, generator or analysis level data can be stored in HDF5 directly without relying on ROOT serialization and ongoing studies are being undertaken by the DUNE and the HEP-CCE group. 

\subsubsection*{Lossy compression of derived data}
HEP experiments routinely are using loss-less compression for their data to reduce storage requirements. For small parts of their data models, some experiments also deploy modules that reduce the precision of data in a very controlled manner. Some experiments have also incorporated lossy compression in specific cases ({e.g.}, IceCube) with success, but currently major experiments do not use more generic methods of lossy compression, despite the fact that the storage precision of derived data can exceed the quality of the measurement by many orders of magnitude. Given fixed storage budgets for future experiments, storing data loss-less with too high precision will mean that not as many records can be stored as for a scenario where lossy compression would limit insignificant bits, so, {e.g.}, trigger rates have to be controlled more strictly.

RAW data is often considered the most valuable outcome of an experiment, at the same time derived data can take up the majority of disk volume. Finding generic methods to limit storage precision of variables can reduce storage required for derived data, without degrading its physics potential or changing the original RAW data.

\subsubsection*{Custom content delivery \& streaming}
Most disk storage is occupied by derived data and detector RAW data requires only limited disk resources. The storage requirements for derived data of multi-purpose collider experiments such as ATLAS and CMS are increased due to multiple physics groups needing different, but overlapping, subsets of the data (different event selection and/or different event content). Both experiments have standard formats, called DAOD-PHYS \& PHYSLITE for ATLAS and nano- \& mini-AOD for CMS, that are intended to satisfy the needs for most analysis and are produced for all events. However, if a physics analysis needs additional information, even for just a sub-sample of the events, alternatives such as adding the content (for all events) to the main data format or producing a separate custom stream are expensive for disk storage. And for the analysis to process on larger/upstream data format can be very slow and burdensome.

A more efficient scenario could include the capability to read additional data on-demand from locations other than the current input file. For example, while processing events in the nano-AOD, a physics analysis workflow may request additional data objects for a sub-sample of events that was stored only in the mini-AOD or was written into a separate location that is accessed only when needed. Such a scenario will need to be supported by a functional persistent navigational infrastructure (as in Ref.~\cite{atlasNavigation}) and robust data streaming capabilities.

\subsection{Storage System Evolution} 

The POSIX IO file system interface was created in 1988 and patterned after the UNIX design principle of ``Everything is a File.'' This design principle made POSIX IO system calls like \texttt{read()} or \texttt{write()} largely independent of the evolution of devices and reduced the need to change APIs whenever new resources became available. However with the advent of hyperscaler providers and open source ecosystems including efficient data formats, in-memory representations, and scalable storage systems, new storage interfaces have emerged. These include Amazon Web Service's S3, object storage, key/value storage, and, more recently, dataset interfaces.

Hyperscaler providers have enough data to design and sufficient clout to order custom storage device hardware from the component industry at very large quantities. However, they need to balance the efficiency and competitive advantage of proprietary designs with the efficiency of open source communities in terms of software development speed and talent development. Through institutions like the Apache Foundation and the Linux Foundations and its many sub foundations, including the TODO Group as the umbrella organization for Open Source Program Offices of nearly a hundred corporations, hyperscalers and their suppliers strategically support open source projects that their businesses depend on. The general availability of the software of these open source projects creates in turn new markets for the component makers that they look to develop without violating their NDAs with hyperscalers. Thus, the HEP community will benefit from aligning their data services stack with open source ecosystems, not only because those communities are large, well-funded, and move quickly, but their designs are indirectly informed by large industry investments in hardware designs that have already proven successful in their proprietary space. One example with potential impact to the HEP computational frontier is computational storage.\footnote{While ``computational storage'' most often refers to computational storage \emph{devices}, we use here the more general notion that includes computational storage \emph{services}.}

\subsubsection*{Computational Storage}

The key advantage of the cloud is its elasticity. This is implemented by systems that can expand and shrink resources quickly and by disaggregation services, including compute and storage. Disaggregation allows compute and storage to scale independently but it places greater demand on expensive top-of-rack networking resources as compute and storage nodes inevitably end up in different racks and even rows as data centers are growing. More network traffic also requires more CPU cycles to be dedicated to sending and receiving data. Network traffic and CPU cycles are key power consumers and can increase latency in case of contention. Therefore, disaggregation, somewhat paradoxically, amplifies the benefit of moving some compute back into storage because data-reducing filtering and compression operations can reduce data movement significantly. 

There is however a catch: in order to move function to data, the \emph{local} data needs to provide the context for the function to succeed: {e.g.,} a projection of a table requires all the metadata and data of that table to be locally available. Many scalable storage systems stripe data across many storage servers or devices to maximize parallel access and workload balance. This striping is based on fixed byte offsets optimized for low-level memory allocation and device controllers instead of semantic completeness of a higher-level data structure such as a table. Thus, striping will have to be semantically aligned instead of byte-aligned. Another name for this is \emph{record-based} storage. Common examples of record-based storage are key/value and object storage (as long as values and objects are not themselves striped) as well as the much older Virtual Storage Access Method (VSAM) commonly used in IBM mainframe applications that pairs files with inverted indices. 

Once semantic partitioning of the data is established, computational storage needs to be able to execute functions on that data. This raises the question of how these functions are implemented and how they are executed. 
For computational storage to be sustainable, data access libraries will have to be able to evolve independently from computational storage, {i.e.}, computational storage will have to support the embedding of access libraries with minimal change. As an example, the IRIS-HEP Skyhook Data Management project recently merged a Ceph plug-in with the Apache Arrow project~\cite{chakraborty:ccgrid22}. The execution of access functions, especially when performed in devices, will need to be sandboxed in some way. Popular technologies are eBPF (ebpf.io) and WebAssembly.

The key technologies to watch out for in the next 5--10 years are: 

\begin{itemize}
  \item Access libraries that map a dataset to multiple kinds of data sources, including multiple instances of itself. This will allow access library instances on clients to push down operations to access library instances embedded in the storage layer. For HEP a particularly interesting example of this because of its 1:1 mapping to Awkward Array, is the Apache Arrow Dataset Interface. These data source abstractions are akin to ``foreign data sources'' that have been popular in relational database access libraries for some time.

  \item Cross-language specification for data compute operations ({e.g.}, substrait.io) which will enable fusing cached query results from multiple data sources into results of new queries.
  
  \item Sandboxing technologies eBPF and WebAssembly with extensions that go beyond what is allowed in the Linux kernel, including floating point calculations, and that can run access libraries efficiently

  \item Distributed resource management for computational storage that dynamically places data management functions balancing locality with occupancy while reducing overall data movement and providing latency guarantees.

  \item Storage devices with computational power and I/O accelerators (for compression and serialization) similar to a smartphone.

\end{itemize}

%% file: Computation/CompF04/topic_analysis_facilities.tex
\section{Analysis Facilities}
\label{sec:af}

\subsection{Definition of Future Analysis Facility}

We define a future analysis facility (AF) as:

\begin{quote}
The infrastructure and services that provide integrated data, software and computational resources to execute one or more elements of an analysis workflow. These resources are shared among members of a virtual organization and supported by that organization.
\end{quote}

This definition is intentionally more broad than the traditional thinking in HEP of end-user analysis facilities as primarily large compute and data centers. These large analysis facilities are important components to support HEP science since they accommodate a large number of use cases and benefit greatly in terms of system management from economy of scale. They also often provide a platform for login access to interactive computing and access to a batch system. Future analysis facilities should also integrate systems and services that are tailored and optimized for specific elements of an analysis workflow. Examples include facilities that provide services for parameter fitting and statistical inference, a system optimized for the training of machine learning models and a Jupyter notebook hosting service. These future systems might support multiple analysis services, or may be single purpose given specific aspects of the hardware platform. Modern analysis tools make a distinct break from the past in that they are often massively parallel and make use of distributed services to operate efficiently and quickly.

Two other important aspects of the future analysis facilities definition are related to sharing and support. For example, while an individual's laptop or desktop might be a crucial part of their analysis infrastructure ({e.g.}, for terminal access to facilities or generating plots), it is not a resource shared broadly with a virtual organization nor is it supported by that organization. What elevates a resource to the level of an analysis facility is official support as a shared resource within an organization of people with shared interests ({e.g.}, a scientific collaboration).
The sharing and support go hand-in-hand, thus making sure a facility is leveraged by the whole community.

\subsection{Challenges Exemplified by the Energy Frontier}

To illustrate some of the specific challenges in designing future Analysis Facilities we discuss ongoing work towards these facilities for the Energy Frontier. We understand these challenges to be exemplary for other areas in HEP as well.

Motivated by the need to probe increasingly rare physics processes, HL-LHC will deliver luminosity to the experiments at roughly an order of magnitude higher rate than previously. For this reason new techniques and services are expected to be used by HL-LHC analysis teams. An Analysis Facility will provide these at scale. To prepare for this new era, prototype facilities must be built, rapidly iterated on, and tested. Testing will need to involve all aspects of a facility from throughput, to ease of use, time-to-insight, and support load.

As part of the integration strategy of software components for analyzing the data as well as the deployment of the analysis software at analysis facilities, IRIS-HEP is organizing an ``Analysis Grand Challenge" \cite{agc}, with a goal to demonstrate and test new technologies envisioned for HL-LHC, including new user interfaces, innovative data access services that provide quick access to the experiment’s data sets, new systems and frameworks such as the Coffea analysis framework \cite{smith2020coffea} allowing analysts to process entire datasets with integrated statistical model building and fitting tools.
Analysis workflows selected for Analysis Grand Challenge include packages and services that also support the reinterpretation and analysis preservation steps allowing to provide long term re-usability of the entire analysis workflow in the future.

The Analysis Grand Challenge workflow defines an analysis benchmark that could be easily re-implemented and executed on any generic Analysis Facility and is designed to help showcase to physicists how to use an existing analysis facility at scale for their analysis.

As evidenced by the response to various workshops, Analysis Facilities topics have been popular across the US LHC community. A number of new approaches and technologies and resource opportunities are now available for analysis facilities. In several submitted Snowmass whitepapers \cite{benjamin2022analysis}, \cite{flechas2022collaborative},\cite{lannon2022analysis}, a set of recommendations and suggestions were developed for analysis facilities with proposed features that could be interesting for both new users and resource providers from an AF development's point of view.

From the resource provision point of view, large HPC centers already are offering opportunities to do large-scale processing. To date, the community has done well adapting these resources for production workloads ({e.g.}, simulation and reconstruction). But they have not been well used for individual user analysis.

There are now modern analysis tools that are being adapted to run on HPC systems (large scale machine-learning, fitting as a service~\cite{Feickert:2021sua}, dataset skimming) that could be available for analysts. This step will require integration with a portal that allows broad access to the community. In other disciplines this is referred to as ``Science Gateways". Such gateways that support tens to hundreds of thousands of researchers exist today for Neuroscience, Genomics, Nanoengineering, and many other disciplines. 
There are thus many existence proofs at scales far exceeding even the largest HEP experiments. These Science Gateways may also be useful as templates for making public HEP data more easily accessible to the wider scientific community beyond the collaborations that produce the data. 

Another important requirement is that facilities must integrate with the existing distributed computing infrastructure, meaning that future analysis facilities will be successful only if they leverage the larger operations and national-scale resource investments that currently exist or are planned. Introducing the new services in analysis facilities ecosystem should be balanced by the important requirement to tightly integrate existing services such as batch systems.

From an infrastructure point of view, the LHC community provides various facilities that serve the independent needs of each experiment. We believe that we need to focus on providing common approaches such as deployment via Kubernetes \cite{k8s} to help exchange ideas at the infrastructure level. Such common approaches have the added value that they are widely used across commercial clouds and academia. It is maybe worth noting that there are roughly 4,000 institutions in the USA engaged in open research, and only three DOE leadership class computing centers. Choosing a common approach as basis for analysis facilities may allow integration of a much larger set of resources across the wider community, reducing cost while increasing access for the community at the same time.

Another crucial R\&D topic for new facilities is to investigate the use of “federated identity” and authorization \cite{bockelman2020wlcg}, \cite{balcas2017cms}. The challenges in this area are as much ``social policy" as ``technical implementation". Crudely speaking, the larger the computing and data facility the more restrictive the security controls and policies tend to be, with the interesting exception of commercial clouds. The latter provide access to vast resources with the main security concern being that the customer proves they are capable of paying their bill.
We propose to investigate the federated identity providers ({e.g.}, new WLCG identity providers) for both web-based and SSH-based access to facilities to facilitate both science gateway and traditional user login via unix account access.

Current analysis facilities are mostly providing the login account and access to disk storage hosting experiment data sets in addition to access to computing resources through a batch system. Jupyterhub \cite{jh} integration with various batch systems should be investigated to improve the user experience.

Another key feature for analysis facilities is the provisioning of authoring and sharing environments, allowing users to easily share their software environments within their own groups and with other groups. The use of containers also greatly improves portability of software and repeatability of environments. 

Another possible R\&D area is the integration of data access services to reduce local storage needs at analysis facilities. This could be achieved by filtering and projecting input data and caching results, which will remove the need for manual bookkeeping for analysts. Investigating object-storage for analysis facilities, which is widely used in industry, should become another priority in the HEP community.

Considering the fact that analysis facilities are specific to a given experiment or community and the software tools are often the same, user support personnel often spend most of their time answering generic support requests. Using an alternate model where experiments could share personnel at shared facilities would allow for lower overall costs, which would be a more sustainable approach.

All the requirements mentioned above for analysis facilities apply equally to the Energy, Neutrino, and other experimental Frontiers.
In this spirit, it will be important for HEP members across all subfields to communicate their developments on analysis facilities. As a current example, the HEP Software Foundation Analysis Facilities Forum~\cite{HSFAF} provides a community platform for those interested in contributing to the development of analysis facilities for use by HEP experiments, serving as a space to develop and exchange ideas.

\subsection{Analysis Frameworks and Integration in Analysis Facilities}

The submitted whitepaper \cite{lannon2022analysis}, strongly emphasized an idea that integration of both hardware and software will be a priority, using the term “analysis cyberinfrastructure” rather than the more common “analysis facility”.

In this section we expand on this notion of “analysis cyberinfrastructure” and how it integrates ``analysis frameworks" and other components as the layers of software stack in an Analysis Facility. 

First, starting from top to bottom, is the \textit{Analysis Software} layer, which includes the software the analysts write themselves along with any direct dependencies. It may take the form of one or more user created libraries and one or more applications built on top of those libraries. In addition, libraries from the broader scientific python ecosystem outside of HEP may be used as integral parts of these applications. 

Defining the \textit{analysis framework} software stack layer, this layer sits directly underneath the analysis software layer to facilitate the interaction between the analysis software and the common reduced data format as well as providing some connection to scale-out mechanisms.

The \textit{user interface} layer is responsible for providing users with the ability to interact with the computational resources.

The \textit{batch infrastructure} layer connects the distributed application and framework to the computational resources to perform the analysis. While the \textit{storage infrastructure} layer consists of a stack with potentially several layers, including the possibility to transform data from one format to another and potentially also caching the results of that transformation.

From the cyberinfrastructure perspective, a focus on integration across the layers can provide important advantages when trying to diagnose potential bottlenecks.
Another open question to consider when exploring approaches to scaling up analysis cyberinfrastructure to support multiple groups is whether it makes sense to focus on the approach of scaling up through a specially-designed physical facility to support multiple analysis groups, to approach analysis cyberinfrastructure as an add-on feature to existing computational facilities (such as LHC Tier-2 sites), or to design analysis cyberinfrastructure to be deployed in a cloud-like approach across a variety of different physical resources from experiment owned sites like LHC Tier-1, 2, and even 3 facilities, to resources that are used on a more temporary basis, such as HPC facilities or even commercial cloud resources.

\subsection{Reproducible Computing Environments and Infrastructure as Code for Analysis Facilities}

The analysis facilities discussed in Ref.~\cite{flechas2022collaborative} were designed from the start to use a container-based infrastructure. Containers provide flexibility, portability and isolation without the additional overhead of virtual machines. Sites that deploy this infrastructure widely make it easier to add elasticity to the analysis facility; servers for a different purpose ({e.g.}, batch worker nodes) can be re-provisioned on the fly for scheduling analysis tasks. The orchestration tool of choice for containers is Kubernetes \cite{k8s}. It provides a unified declarative description and configuration language, configuration management, service discovery, service load balancing, automated roll-outs and rollbacks, and other features key to providing stable services.

Kubernetes was originally designed for cloud computing, which adopts a single-tenant model: one user creates and owns an entire cluster. Since its original public release, it has been extended with much more complex base-access controls, policy primitives, and a configurable programmable filter module in front of the API. For this reason pure Kubernetes is a good fit for facilities such as Coffea-casa \cite{adamec2021coffea} at University Nebraska-Lincoln which are designed to serve a single experiment. For multi-tenant facilities there exists Red Hat’s open-source OKD platform \cite{okd}, which is a super-set of Kubernetes. OKD incorporates additional security and isolation, adds operations-centric tools, a user friendly GUI, and additional storage and network orchestration components, making OKD a good choice for the Elastic Analysis Facility at Fermilab or any other DOE funded analysis facility.

Analysis facilities, such as Coffea-casa \cite{adamec2021coffea} and Elastic Analysis Facility in Fermilab build and use custom containers to facilitate the integration of such a complex application. Although the team builds and maintains these images, the versatility of Kubernetes \cite{k8s} allows for the drop-in replacement of other custom Jupyter notebook containers, and the Jupyterhub \cite{jh} instance can be configured to allow user selection of supported images.

Another more sustainable option to provide more flexible reproducible solution is a Binderhub \cite{ragan2018binder}, which is a Kubernetes-based cloud service that can launch a repository of code (from GitHub, GitLab, and others) in a browser window such that the code can be executed and interacted with in the form of a Jupyter notebook. Binder, the product behind mybinder.org as a user interface, is also useful for reproducibility because the code needs to be version controlled and the computational environment needs to be documented in order to benefit from the functionality of Binder.

From an infrastructure point of view, using Infrastructure as Code (IaC) \cite{morris2016infrastructure} offers advantages for auditing and reproducibility. GitOps \cite{beetz2021gitops} is defined as a model for operating Kubernetes clusters or cloud-native applications using the version control system Git as the single source of truth. One of the features GitOps envisions is declarative descriptions of an environment to be stored as infrastructure- as-code in a Git repository. Continuous integration, delivery, and deployment are software development industry practices that enable organizations to frequently and reliably release new features and products. They allow for rapid collaboration, better quality control, and automation as code goes through formal review processes and is audited and tested on a regular basis.

\subsection{Analysis Facility Summary: Future Work}

We have described a set of requirements we believe that could be considered 
for upcoming years for the Analysis Facilities architects and developers. 

We suggest to work on prototyping of Analysis Facilities together with analysis software developers, resource providers and analysis facility architects using modern techniques, such as exploring concepts of ``Infrastructure as Code'', the integration of ``federated identities", to facilitate the preservation of user environments and many others. 

We also recommend that Analysis facilities  be made interoperable, allowing users  to navigate seamlessly from one Analysis Facility to another, and easily extensible, to accommodate future needs without disruptions.

%% file: Computation/CompF04/topic_edge_services.tex
\section{Edge Services}
\label{sec:edge}

\subsection{Definition}

We adopted the following definition for edge services: 

\begin{quote}
Edge Services operate at the interface between a data center and the wide area network, separated from the data center's core services.
This includes middleware that facilitates user access between the data center and external systems ({e.g.}, storage, databases, workflow managers).
These services may be managed externally in partnership with the data center and federated across multiple data centers.
\end{quote}

Here, ``wide area network'' can mean the Internet as a whole, but could also include specialized network connections for specific services.
Edge Services may also serve applications on the ``interior'' of a data center, such as a bespoke workflow manager or database.
We also note that these services could be user-supplied, and run on infrastructure that is itself considered an Edge Service. Finally, we note
that definitions are evolving rapidly and may diverge between different disciplines. For example, approximately five years ago, the term
would have primarily referred to data caching services related to data delivery.  Outside the HPC community, ``Edge Services'' may sometimes be associated
with the ``Internet of Things (IoT)''.  Therefore, while useful, our definition should not be regarded as exclusive or permanent. 

The current and proposed activity in ``Analysis Facilities'', as covered in Section \ref{sec:af}, makes heavy use of these edge services and so some of the research activity covered here overlaps with that proposed in that section though it has broader applicability. 

\subsection{Recurring Themes}

There are many examples of existing Edge Services, but to focus on the
future, we need to look at the common features of these services.

Many modern services are based fundamentally around the idea of containers: images
that contain all the software and library dependencies that are needed to
execute a particular application.  These containers can be assembled in
various ways to form a full application stack.  Most commonly Kubernetes~\cite{k8s} is now
used to deploy these applications, though there are alternatives.

These applications are typically configured using a declarative language
that can be stored in a version-controlled repository, such as GitHub, and
then deployed using automated tools.  In the wider technology industry, this
is called ``DevOps''.  Other terms that capture the same idea are
``System as software'' or ``Infrastructure-as-a-Service'' (see also Section~\ref{sec:af}).

Since these services are easily deployable by design, we have seen that
``federated'' services are becoming increasingly common.  The same
application may be active on many different data centers, and there is 
a closely related concept of ``federated identity'', where authentication
provided by one laboratory or data center can be shared among many data 
centers.

\subsection{Concerns}

The common themes described above address many of the concerns that have been raised by the community around this topic.  For example, 
systems originally developed in the c.2000--2010 era were highly
specialized and are no longer affordably maintainable.  By leveraging
open-source and open-infrastructure capabilities instead, future teams can
easily pick up and deploy edge service tools.

As mentioned above, many edge services are deployed using Kubernetes.  While
powerful, this application is known to be difficult to learn, configure
and maintain.  Simpler management tools are needed to assist the process
of migrating applications to this system.

Security is, of course, a large, if not the largest, concern. Indeed, the concept
of federated services also demands federated security.  We have seen an
increasing reliance on third-party authentication, such as OpenID~\cite{OpenID}, Google Identity~\cite{GoogleID},
or ORCID~\cite{ORCID}, among many examples.  While this does reduce workload on individual
data centers, who do not need to deploy and maintain their own authentication
mechanisms, there is a risk that these services could become unavailable,
and centers need to be prepared to at least alert end-users when, for example,
ORCID is not a valid authentication mechanism because it is temporarily
offline.

\subsection{Future Needs}

Looking forward to the next decade, we can already see situations where the needs
of scientific applications, even for smaller HEP collaborations and projects, are at a scale beyond the capacity of individual
data centers.  This is where federated services will be in high demand.
In fact, individual data centers may have different internal policies and procedures,
but federated edge services can ``glue'' those centers together and provide
a unified application to enable science.

We also foresee an increasing number of user-supplied applications that enable
domain-specific or project-specific science.  For example, the Spin~\cite{Spin} service at
NERSC~\cite{NERSC} allows any NERSC user to create services that access their own data.  In
effect, this edge service enables the creation of other edge services.
This trend should be encouraged, and the developers of the underlying 
infrastructure should be aware of demand for ``push-button'' services.
For example, individual users should be able to create a database service
with minimal effort, given a template provided by a data center providing
such a service.

Data center and HPC center managers need to be aware of this trend. Although many
centers are working on provisioning edge services, these services still
do not necessarily rise to the attention of the highest level of planning. We
believe that immediate and near-future efforts should be directed to address
this concern.

Finally, we note that the edge services we have examined enable science
well beyond the field of high-energy physics. For example, we have examined
edge service applications relevant to astrophysics, genomics and microbiology
among many other fields.  As part of the Snowmass process
and beyond, the high-energy physics community should reach out to and share
experience with other disciplines.

%% file: Computation/CompF04/topic_networking.tex
\section{Networking}

\label{sec:net}

Networking, in conjunction with computing and storage, are key enablers for all aspects of Particle Physics - experiments, data analysis, and discovery. Particle Physics has a strategic commitment to custodial responsibility for experimental and observational data sets. The computing and data analysis landscape is in a state of continuous evolution and change, with the location and technology of data analysis moving as technology evolves. Networking is the data circulatory system for scientific collaborations, transporting science data (the “crown jewels” of the science community) to computing and data analysis, and the results back to the collaboration. However unlike compute and storage resources, the perception of networking is that it is ubiquitous, unlimited, and unpredictable.  This perception, supported by many years of exponential capacity growth, will need to evolve during the next decade.   We foresee two reasons for this: 1) the exponential increases in capacity for fixed cost will begin to see hard limits over this period and 2) the globalization of many other science domains will significantly increase the demands on research and education networks.  {\bf Therefore, we believe that it is critical to take steps now to ensure we will have the network capacity and capability required to effectively pursue particle physics science goals over the next decade and beyond.}
\subsection{Overview and Motivation}
Research and Education (R\&E) network traffic continues to grow at an exponential rate~\cite{MyESnet}, with traffic from Particle Physics expected to grow by a factor of 10 between 2022 and 2029~\cite{ESnet-HEP-Requirements}.  Historically, network capacity and technology upgrades have kept up with demand, however, we are quickly approaching the physical limits ({e.g.}, Shannon’s limit~\cite{Shannon-Ciena,Shannon-HuaWei}, coherent detection~\cite{CoherentDetectLimit}) of the advancements that can be made, and simply deploying more physical infrastructure ({e.g.}, laying down new fiber) may prove cost prohibitive.  In addition, the use of the network as an unpredictable “black-box” resource results in significant inefficiencies in today’s complex and widely distributed workflows. To this end, we propose the following four areas of research and development that would enable more interactive, intelligent, and fair use of network resources moving forward.
\begin{itemize}
\item {\bf Network Interaction Optimization} --- Capabilities or functions that allow the application to better interact with the network, resulting in improved performance or enhanced features.  Examples of such activities include network traffic shaping and packet pacing (e.g., Linux tc~\cite{LinuxTC}, or the behavior of BBR TCP~\cite{10.1145/3012426.3022184,bbrv2}), transitioning to IPv6~\cite{M-21-07}, and multi-domain source based routing~\cite{SegRouting,SegRtHead}.
\item {\bf Resource Orchestration and Automation} --- The ability to intelligently coordinate the  scheduling and provisioning of network resources to facilitate predictable data movement behaviors.  Examples of such activities include site traffic steering~\cite{NOTED-FTS}, network and DTN resource orchestration~\cite{Whitepaper-DataTrans-NetServices}, compute APIs~\cite{SuperFacilityAPI}, white-box switches and SDN routing~\cite{RARE}, AI/ML driven network utilization prediction and traffic engineering~\cite{HECATE}, as well as frameworks for integrated facilities~\cite{DOE-IRI-whitepaper}.
\item {\bf Network and Traffic Visibility} ---  Insight into network health and traffic flow patterns to guide data movement decisions, and direct troubleshooting efforts.  Examples of such activities include  packet marking and flow labeling~\cite{RNTWG-firefly}, and high fidelity network telemetry~\cite{Kim2015InbandNT,10.1145/3452411.3464443},
\item {\bf Data Movement Optimization} --- Capabilities or functions that can improve the end user experience by reducing the time to fetch data. Data Movement Optimization is a key component of improving network performance while coping with significantly higher future demand within the limitations imposed by finite funding resources. Examples of such activities include in-network caching~\cite{NetCacheDistScience} to reduce the latency of data access and reduce long-haul network traffic load, and multi-domain traffic load-balancing~\cite{TransLayerNet} to reduce traffic congestion and increase throughput and resiliency.
\end{itemize}
While the proposed areas above are considered discrete, taking a systems approach is key. Optimization of individual components will not result in an integrated workable system, as there are too many inter-dependencies, and many of the capabilities that the community needs are emergent properties of synergistic interaction between multiple components of a larger system. Having a system with these capabilities will allow Particle Physics to make better networking decisions, increase workflow predictability, and potentially reduce the overall time to results.
 
It is imperative to understand that networking is an end-to-end service which involves multiple domains, and as such, requires collaborating networks to provide inter-operable and congruent capabilities, as well as usage policies that are aligned. Equally important is the notion that security must be an integral part of the research, prototyping, and production implementation, and not an afterthought.

Lastly, it is well understood that the technology landscape changes quickly over time, and well-organized collaborations staffed with knowledgeable experts can make effective use of current and future technologies, whatever they may be. It is critical that Particle Physics make long-term investments in collaborations between scientists and technologists so that cutting-edge networking technologies can be effectively used by Particle Physics. These collaborations must combine research, prototyping and production implementation, as this is the only way that components and technologies can be effectively integrated into the scientific enterprise as effective capabilities. 

\subsection{Network Interaction Optimization}
There are multiple interaction points between a network application ({e.g.}, a data transfer service such as FTS) and the network itself. Optimizing these capabilities can have significant benefits, including increased application performance, increased application flexibility, increased network efficiency, and deterministic path selection for reliability. Optimizations of this kind represent a set of incremental improvements, which can result in significant improvements when taken together. Several examples of current optimization efforts are described here to illustrate the breadth of options available, but the key point is that as technology evolves it will remain important to have experts working on Network Interaction Optimization so as to be able to continuously improve the interaction between Particle Physics applications and the networks that interconnect them. Network traffic shaping can reduce the burstiness commonly associated with TCP’s interaction with the wide area network, resulting in less traffic variability and fewer instances of packet loss. Traffic pacing is a similar capability, which can be configured in the Linux kernel using tc~\cite{LinuxTC} or incorporated into protocols as exemplified by TCP BBR~\cite{10.1145/3012426.3022184,bbrv2}. Beyond the behavior of individual packets or specific protocols, science networks are increasingly able to provide network paths or channels with specific capabilities to scientific applications - including traffic engineering (guaranteed bandwidth and explicit path). This can be accomplished using services such as OSCARS~\cite{OSCARS} or by deploying segment routing~\cite{SegRouting,SegRtHead} in the network. In addition to optimizing protocol and packet interactions, it is important that the set of organizations and entities (networks, sites, caches, computing systems, storage systems) interact using an interoperable and coherent set of mechanisms. Coordination across many infrastructure-aware tools and systems is a capability in itself, which requires research, prototyping, and transition to production. These interactions are complex, and deserve their own research and development effort.

\subsection{Resource Orchestration and Automation}
Domain science workflows are currently forced to view the network as an opaque infrastructure into which they inject data and hope that it emerges at the destination with an acceptable Quality of Experience.  There is little ability for applications to interact with the network to exchange information, negotiate performance parameters, discover expected performance metrics, or receive status/troubleshooting information in real time.   Resource orchestration which includes the network along with the compute, storage, and instrument systems will be needed as the trend toward large team distributed collaborations increases.  This orchestration of workflow dependent network resources will allow deterministic network performance around which science workflows can plan and adjust.  Software driven network control has matured to the point where it can be applied in service of domain science workflow objectives.  Leveraging AI/ML innovations to predict usage and help drive resource allocation decisions will also need these integrated orchestration mechanisms to fully realize system optimizations.  These types of network focused orchestration and automation technologies are identified as key enabling technologies to realize the DOE Integrated Facilities vision as outlined in Ref.~\cite{DOE-IRI-whitepaper}.   Some of the key considerations and technologies challenges include:  i) API and AuthN standardization and/or other mechanisms to simplify access to orchestrated services, ii) ease of workflow use will require sophisticated network side monitoring and troubleshooting functions,  iii) AI based network control systems which are verifiable, monitorable, and controllable.  

There are multiple ongoing  projects focused on these types of network resource automation and integrated orchestration technologies.   The NOTED~\cite{NOTED-FTS} project is building workflow specific network use optimization tools.  The SENSE~\cite{SENSE} project has developed a multi-resource multi-domain orchestration system.  The NERSC developed Superfacility API~\cite{SuperFacilityAPI} enables automated HPC usage. The RARE~\cite{RARE} project is focused on programmable network dataplanes.   Network system focused AI/ML projects include the HECATE~\cite{HECATE} project which is developing self-driving network technologies.

\subsection{Network and Traffic Visibility}
The ability to view the status of compute jobs is fundamental to understanding how an analysis process is progressing.  Unfortunately this level of transparency is typically unavailable for networking resources, and as such, networking is typically perceived as a “black-box”. Precision network telemetry and high-fidelity traffic flow tracking can provide unprecedented insight into network health and traffic movement patterns, and drive informed decision making.

Precision network telemetry information that is accessible to applications (in real-time) can be extremely valuable in setting expectations, understanding performance issues, and guiding intelligent decisions on when data movements should be scheduled. Technology solutions such as the P4 In-band Network Telemetry~\cite{Kim2015InbandNT} and the ESnet High-Touch platform~\cite{10.1145/3452411.3464443} can provide real-time per packet information of how a data flow is performing. Additionally by observing the flow at different points in the network, it is possible to pin-point the locality of network performance issues.

High fidelity traffic flow tracking is important for accurate data movement analysis and auditing, and developing precise usage models. Understanding how related data sets transits a network can provide invaluable insight into capacity planning and traffic engineering decisions. This is especially important where bandwidth is comparatively constrained, such as the trans-oceanic links. Activities such as the RNTWG~\cite{RNTWG} packet marking~\cite{RNTWG-packetmarking,RNTWG-firefly} is an example of large scale flow tracking analysis, spanning multiple network domains.

Networking, by its nature, relies on a richly connected fabric of network providers. This has two obvious implications as it pertains to network telemetry and traffic flow information. Firstly, the information is only useful if it is being collected, {i.e.}, instrumented across the various networks, and secondly, if the information can be shared. Having a unified statistics platform across WLCG sites would go far to facilitate end-to-end multi-domain traffic analysis. Additionally, a common AuthN framework with bilateral trust relations would be beneficial if sensitive data are to be accessed.

\subsection{Data Movement Optimization}
Most, if not all networks, operate under the assumption of best-effort delivery.  This is the result of statically configured link metrics that are used to determine the “best” path between the network ingress and egress. Such practices often lead to unmitigated transient congestion and inefficient use of the network. Techniques such as in-network caching, multi-path end-to-end load-balancing, and meta-scheduling, can be utilized to reduce the inadequacies of best-effort delivery.

In-network caching can reduce the time to retrieve data and improve workflow performance.  This is especially true if the placement of the data is geographically local to the receiver.  An added benefit to in-network caching is that it can be used in conjunction with scheduling algorithms to reduce traffic congestion in the network.  Efforts such as the OSG in-network caching pilot~\cite{NetCacheDistScience} and WLCG data lakes~\cite{WLCG-DataLakes} demonstrate the benefit of a network caching model.  From a deployment standpoint, it should be noted that administrating “3rd party” caching stacks requires a non-trivial amount of coordination ({e.g.}, acquiring the appropriate certificates, balancing domain security concerns for access, negotiating support models, etc.) for on-going operations.

Multi-path end-to-end load-balancing allows for several benefits, such as alleviating hot-spots in the network, using underutilized network paths, and enhancing application level data transfer resiliency. To effectively perform load-balancing above the network layer~\cite{OSI-Model} at high speeds, it requires hardware that can take session layer information to determine which data packets constitute the same flow, and steer the data packets over different paths~\cite{TransLayerNet}. The granularity of how load-balancing is executed can range from selecting different network domains in the end-to-end path, down to specific paths within a network. In both cases, an understanding of network routing policies is necessary, along with some method to interact with the network (see “Network Interaction Optimization” and “Resource Orchestration and Automation” sections above). With load-balancing being performed above the network layer, there is a requirement that both the source and destination ends must possess the same capability to ensure proper segmentation and reassembly of the load-balanced data flow.

Meta-scheduling is a complementary approach to existing and new traffic engineering mechanisms that can make efficient use of available network capacity through job awareness, keeping key components of network infrastructure, such as trans-atlantic links, cost effective.  Analogous to how existing workload management systems consider computational aspects such as cores and available storage when scheduling a job, a meta-scheduler that is network aware could manage network resources in a similar manner. For example, TEMPUS~\cite{kandula2014calendaring} manages the scheduling of both long-running and high priority transfers while considering the economic models of cloud resources, and Pretium~\cite{jalaparti2016dynamic} uses pricing models to drive traffic engineering decisions.  Other approaches such as DIANA~\cite{mcclatchey2007data} demonstrate how meta-scheduling can be integrated into a complex workload management system, and can be extended to leverage data management and software defined networking techniques.

\subsection{Network Summary: Challenges and Needed Work}

We have described a set of four broad areas in networking we believe will need active effort during the coming years.  Noting that technology evolution will change the way the above concepts are implemented, it is critical to have structures in place to track, implement and integrate new technologies.  {\bf Especially important is the transition from research into production, which will require significant effort and should not be underestimated.}

Prototypes offer powerful means of demonstrating new technologies and capabilities, allowing evaluation of cost, complexity, effort, and maintainability.  In all the identified areas, we suggest that there be work plans that clearly identify the steps and decision points from prototyping to production.

Security considerations are critical, starting from initial design all the way to production, and needs to be part of any process that will provide our future infrastructure components.  

Of special consideration for the networking space is accountability and fairness within and between experiments.  If technology allows varying levels of service across the networks, how will that be managed in equitable ways between users of the network, from individual scientists to large collaborations?  

A central aspect of future networks will be their capacity (bandwidth) and its associated cost evolution.  If the world-wide set of science collaborations have global traffic demands evolving faster than what the R\&E networks can afford, we will be driven towards mechanisms that focus on the efficient, equitable management of the available network capacity.  We note that this would be a significant change in the network environment and will require new tools, approaches and mechanisms to operate efficiently.  While there are varying opinions on how likely such scenarios are, we must prepare for them years in advance to realistically expect to have the needed capabilities if such scenarios arise.

%% file: Computation/CompF04/Appendix.tex
\appendix
\section{Workshop Participants} \label{app:participants}
As well as submitted whitepapers,  this report is the result of community discussions, including sessions in the Computational Frontier workshop (August 10--11, 2020) \url{https://indico.fnal.gov/event/43829/timetable/#20200810} and the CompF4 Topical Group workshop (April 7--8, 2022) \url{https://indico.fnal.gov/event/53251/}. The registered paripants of those workshops are listed below (only those who selected CompF4 are listed for the August 2020 workshop).
\clearpage
\subsection*{Computational Frontier Workshop (August 10-11, 2020) Participants}
\includegraphics[width=1.1\textwidth,page=1]{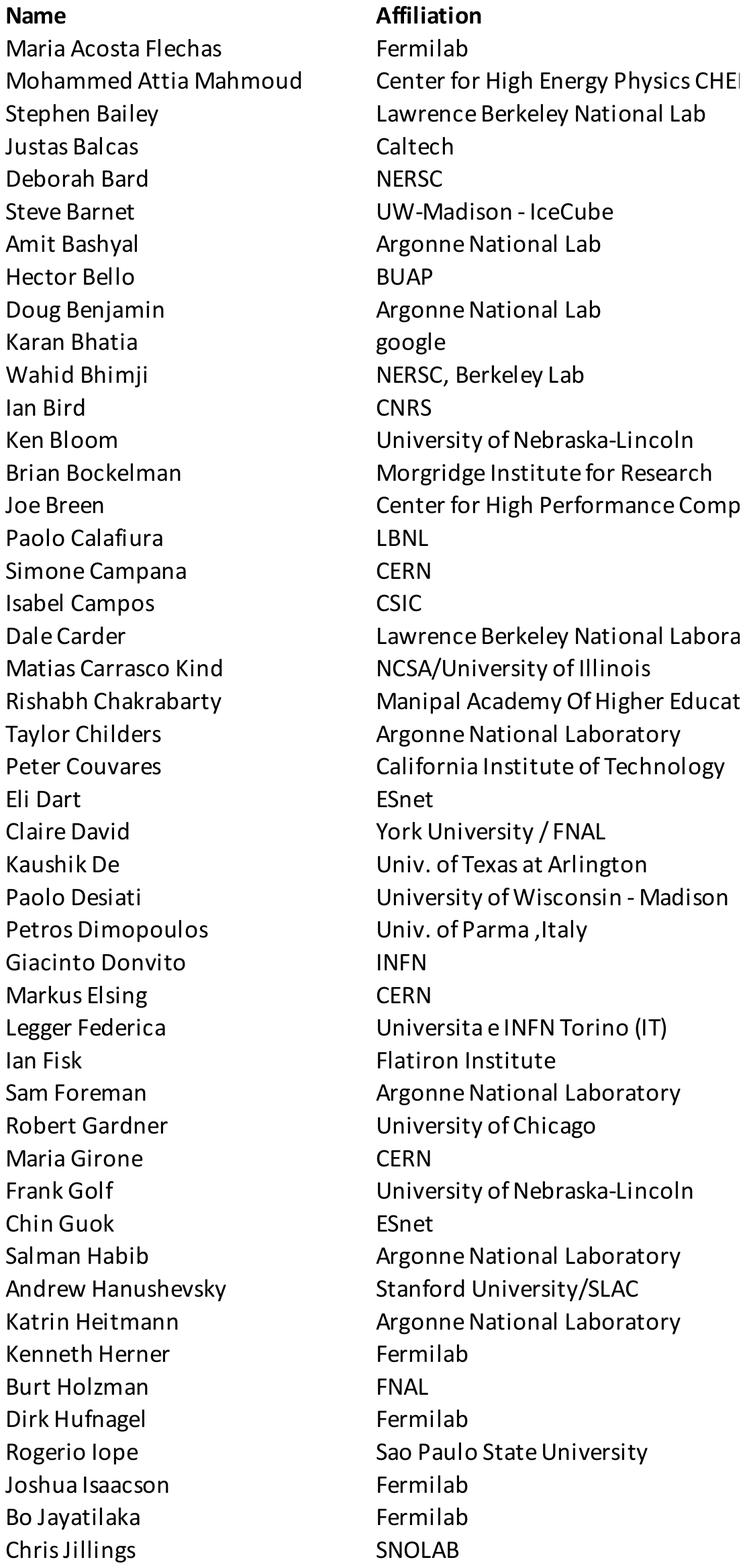}\clearpage 
\includegraphics[width=1.1\textwidth,page=2]{Computation/CompF04/figures/Registration-CompF4-Aug10-11-2020.pdf} \clearpage 

\includegraphics[width=1.1\textwidth,page=3]{Computation/CompF04/figures/Registration-CompF4-Aug10-11-2020.pdf}


\subsection*{Comp4 Workshop (April 7-8, 2022) Participants}
\includegraphics[width=1.1\textwidth,page=1]{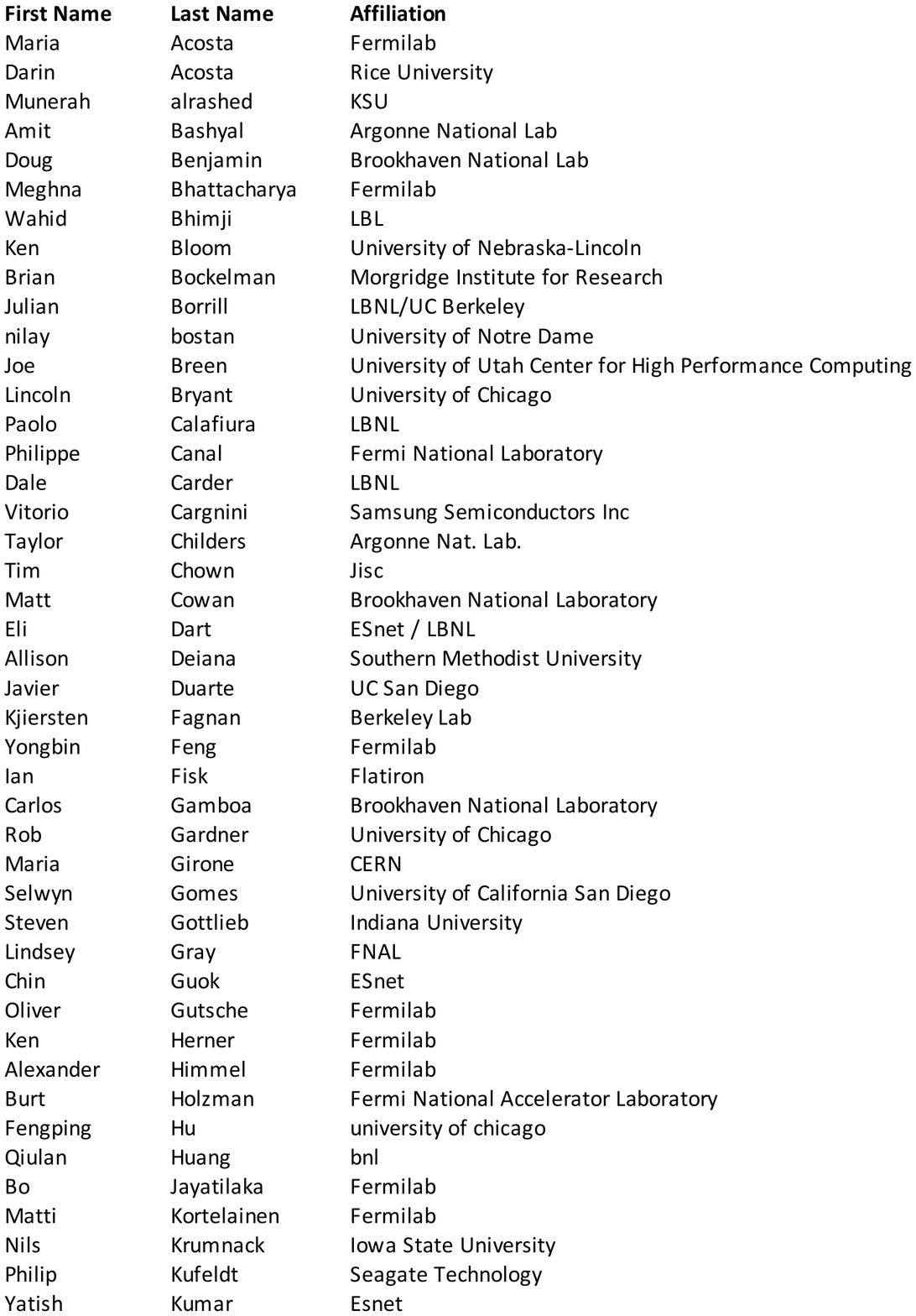}
\clearpage
\includegraphics[width=1.1\textwidth,page=2]{Computation/CompF04/figures/CompF4_April2022_RegistrationReport.pdf}
\clearpage 

\section{Topic Leads} \label{app:topicleads}
The co-conveners for this topical group were Wahid Bhimji, Rob Gardner (Until Fall 2021), Meifeng Lin (From Fall 2021) and Frank W\"urthwein. In addition, the topic leads in the table below heavily contributed to this process, including soliciting and reviewing whitepapers, coordinating sessions in the April 2022 workshop, bringing together other materials from the community, and writing relevant sections of this report.

\begin{table}[h!]
    \centering
    \begin{tabular}{|c|c|}
    \hline
    \textbf{Topic} & \textbf{Leads} \\ 
    \hline \hline 
    Processing & Meifeng Lin and Ian Fisk \\
    \hline
    AI Hardware  & Javier Duarte and Nhan Tran \\
    \hline
    Storage & Carlos Maltzahn, Peter Van Gemmeren and Bo Jayatilaka \\
    \hline
    Analysis facilities & Oksana Shadura, Mark Neubauer and Gordon Watts \\
    \hline
    Edge Services & Ofer Rind and Benjamin Weaver  \\
    \hline
    Networking & Eli Dart, Chin Guok and Shawn McKee \\
    \hline

    \end{tabular}
    \label{tab:my_label}
\end{table}

%% file: arxiv-main.bbl
\providecommand{\href}[2]{#2}\begingroup\raggedright\begin{thebibliography}{10}

\bibitem{Aug2020}
``{Snowmass 2021 Computational Frontier Workshop, August 10-11, 2020}.''
  \url{https://indico.fnal.gov/event/43829/timetable/#20200810}.

\bibitem{April2022}
``{Snowmass 2021 CompF4 Topical Group Workshop, April 7-8, 2022}.''
  \url{https://indico.fnal.gov/event/53251/}.

\bibitem{ai-taskforce}
``{National Artificial Intelligence Research Resource Task Force (NAIRRTF)}.''
  \url{https://www.ai.gov/nairrtf/ }.

\bibitem{LHCComp:2022}
S.~Campana, A.~Di~Girolamo, P.~Laycock, Z.~Marshall, H.~Schellman and
  G.A.~Stewart, \emph{Hep computing collaborations for the challenges of the
  next decade},  2022.
\newblock 10.48550/ARXIV.2203.07237.

\bibitem{osti_1581234}
B.~Joó, C.~Jung, N.H.~Christ, W.~Detmold, R.G.~Edwards, M.~Savage et~al.,
  \emph{Status and future perspectives for lattice gauge theory calculations to
  the exascale and beyond},
  \href{https://doi.org/10.1140/epja/i2019-12919-7}{\emph{European Physical
  Journal. A} {\bfseries 55} (2019) }.

\bibitem{Kahn:2022kae}
Y.~Kahn et~al., \emph{{Snowmass2021 Cosmic Frontier: Modeling, statistics,
  simulations, and computing needs for direct dark matter detection}},  in
  \emph{{2022 Snowmass Summer Study}}, 3, 2022
  [\href{https://arxiv.org/abs/2203.07700}{{\ttfamily 2203.07700}}].

\bibitem{lqcd-boyle}
P.~Boyle, D.~Bollweg, R.~Brower, N.~Christ, C.~DeTar, R.~Edwards et~al.,
  \emph{Lattice qcd and the computational frontier},  2022.
\newblock 10.48550/ARXIV.2204.00039.

\bibitem{SmallExp:2022wp}
D.~Casper, M.E.~Monzani, B.~Nachman, C.~Andreopoulos, S.~Bailey, D.~Bard
  et~al., \emph{Software and computing for small hep experiments},  2022.
\newblock 10.48550/ARXIV.2203.07645.

\bibitem{digital_twins}
A.~El~Saddik, \emph{Digital twins: The convergence of multimedia technologies},
  \href{https://doi.org/10.1109/MMUL.2018.023121167}{\emph{IEEE MultiMedia}
  {\bfseries 25} (2018) 87}.

\bibitem{FASER:2022yqp}
{\scshape FASER, ATLAS, LZ, Fermi-LAT, H1, T2K, SBND} collaboration,
  \emph{{Software and Computing for Small HEP Experiments}},  in \emph{{2022
  Snowmass Summer Study}}, D.~Casper, M.E.~Monzani, B.~Nachman and G.~Cerati,
  eds., 3, 2022 [\href{https://arxiv.org/abs/2203.07645}{{\ttfamily
  2203.07645}}].

\bibitem{girone_maria_2020_3647548}
M.~Girone, \emph{{Common challenges for HPC integration into LHC computing}},
  Feb., 2020.
\newblock 10.5281/zenodo.3647548.

\bibitem{Bhattacharya:2022qgj}
M.~Bhattacharya et~al., \emph{{Portability: A Necessary Approach for Future
  Scientific Software}},  in \emph{{2022 Snowmass Summer Study}}, 3, 2022
  [\href{https://arxiv.org/abs/2203.09945}{{\ttfamily 2203.09945}}].

\bibitem{Jones:2022ycw}
C.D.~Jones, K.~Knoepfel, P.~Calafiura, C.~Leggett and V.~Tsulaia,
  \emph{{Evolution of HEP Processing Frameworks}},  in \emph{{2022 Snowmass
  Summer Study}}, 3, 2022 [\href{https://arxiv.org/abs/2203.14345}{{\ttfamily
  2203.14345}}].

\bibitem{Bartoldus:2022zlc}
R.~Bartoldus, C.~Bernius and D.W.~Miller, \emph{{Innovations in trigger and
  data acquisition systems for next-generation physics facilities}},  in
  \emph{{2022 Snowmass Summer Study}}, 3, 2022
  [\href{https://arxiv.org/abs/2203.07620}{{\ttfamily 2203.07620}}].

\bibitem{10.3389/fdata.2022.787421}
A.M.~Deiana, N.~Tran, J.~Agar, M.~Blott, G.~Di~Guglielmo, J.~Duarte et~al.,
  \emph{Applications and techniques for fast machine learning in science},
  \href{https://doi.org/10.3389/fdata.2022.787421}{\emph{Frontiers in Big Data}
  {\bfseries 5} (2022) }.

\bibitem{a3d3}
{A3D3 Institute}, ``{A3D3 Institute}.'' \url{https://a3d3.ai/}, 2022.

\bibitem{mattson2020mlperf}
P.~Mattson, V.J.~Reddi, C.~Cheng, C.~Coleman, G.~Diamos, D.~Kanter et~al.,
  \emph{{MLPerf: An industry standard benchmark suite for machine learning
  performance}}, {\emph{IEEE Micro} {\bfseries 40} (2020) 8}.

\bibitem{farrell2021mlperf}
S.~Farrell, M.~Emani, J.~Balma, L.~Drescher, A.~Drozd, A.~Fink et~al.,
  \emph{{MLPerf HPC: A Holistic Benchmark Suite for Scientific Machine Learning
  on HPC Systems}},  in \emph{2021 IEEE/ACM Workshop on Machine Learning in
  High Performance Computing Environments (MLHPC)}, p.~33, IEEE, 2021
  [\href{https://arxiv.org/abs/2110.11466}{{\ttfamily 2110.11466}}].

\bibitem{inferencedatacenter2021}
MLCommons, \emph{Inference datacenter v1.1},  2021.

\bibitem{inferenceedge2021}
MLCommons, \emph{Inference edge v1.1},  2021.

\bibitem{reddi2020mlperf}
V.J.~Reddi, D.~Kanter, P.~Mattson, J.~Duke, T.~Nguyen, R.~Chukka et~al.,
  ``{MLPerf Mobile Inference Benchmark}.'' 2020.

\bibitem{banbury2021mlperf}
C.~Banbury, V.J.~Reddi, P.~Torelli, J.~Holleman, N.~Jeffries, C.~Kiraly et~al.,
  \emph{{MLPerf Tiny} benchmark},  in \emph{Proceedings of the Neural
  Information Processing Systems Track on Datasets and Benchmarks}, vol.~1, 12,
  2021,
  \href{https://datasets-benchmarks-proceedings.neurips.cc/paper/2021/hash/da4fb5c6e93e74d3df8527599fa62642-Abstract-round1.html}{https://datasets-benchmarks-proceedings.neurips.cc/paper/2021/hash/da4fb5c6e93e74d3df8527599fa62642-Abstract-round1.html}
  [\href{https://arxiv.org/abs/2106.07597}{{\ttfamily 2106.07597}}].

\bibitem{benchcouncil}
BenchCouncil, \emph{Aibench},  2018.

\bibitem{ignatov2019ai}
A.~Ignatov, R.~Timofte, A.~Kulik, S.~Yang, K.~Wang, F.~Baum et~al., \emph{Ai
  benchmark: All about deep learning on smartphones in 2019},  in \emph{2019
  IEEE/CVF International Conference on Computer Vision Workshop (ICCVW)},
  p.~3617, IEEE, 2019.

\bibitem{torelli2019measuring}
P.~Torelli and M.~Bangale, \emph{Measuring inference performance of
  machine-learning frameworks on edge-class devices with the {MLMark}
  benchmark},  White Paper
  \href{https://www.eembc.org/techlit/articles/MLMARK-WHITEPAPER-FINAL-1.pdf}{}
  (2019).

\bibitem{aimatrix}
{Alibaba}, \emph{{AI matrix}},  2018.

\bibitem{aixprt}
{Principled Technologies}, \emph{{Aixprt community preview}},  2019.

\bibitem{deepbench}
Baidu, \emph{{DeepBench}: Benchmarking deep learning operations on different
  hardware},  2017.

\bibitem{zhu2018benchmarking}
H.~Zhu, M.~Akrout, B.~Zheng, A.~Pelegris, A.~Jayarajan, A.~Phanishayee et~al.,
  \emph{Benchmarking and analyzing deep neural network training},  in
  \emph{2018 IEEE International Symposium on Workload Characterization
  (IISWC)}, p.~88, IEEE, 2018
  [\href{https://arxiv.org/abs/1803.06905}{{\ttfamily 1803.06905}}].

\bibitem{adolf2016fathom}
R.~Adolf, S.~Rama, B.~Reagen, G.-Y.~Wei and D.~Brooks, \emph{Fathom: Reference
  workloads for modern deep learning methods},  in \emph{2016 IEEE
  International Symposium on Workload Characterization (IISWC)}, p.~1, IEEE,
  2016.

\bibitem{james2020rlbench}
S.~James, Z.~Ma, D.R.~Arrojo and A.J.~Davison, \emph{Rlbench: The robot
  learning benchmark \& learning environment}, {\emph{IEEE Robotics and
  Automation Letters} {\bfseries 5} (2020) 3019}.

\bibitem{coleman2017dawnbench}
C.~Coleman, D.~Narayanan, D.~Kang, T.~Zhao, J.~Zhang, L.~Nardi et~al.,
  \emph{Dawnbench: An end-to-end deep learning benchmark and competition},
  {\emph{Training} {\bfseries 100} (2017) 102}.

\bibitem{mlcommonsscience}
MLCommons, \emph{Science working group},  2020.

\bibitem{thiyagalingam2021scientific}
J.~Thiyagalingam, M.~Shankar, G.~Fox and T.~Hey, ``Scientific machine learning
  benchmarks.'' 2021.

\bibitem{toptagging}
G.~Kasieczka, T.~Plehn, A.~Butter, K.~Cranmer, D.~Debnath, B.~Dillon et~al.,
  \emph{The machine learning landscape of top taggers},
  \href{https://doi.org/10.21468/SciPostPhys.7.1.014}{\emph{SciPost Physics}
  {\bfseries 7} (2019) }.

\bibitem{Amrouche:2021tio}
S.~Amrouche et~al., ``{The Tracking Machine Learning challenge : Throughput
  phase}.'' 5, 2021.

\bibitem{Duarte:2019fta}
J.~Duarte et~al., \emph{{FPGA-accelerated machine learning inference as a
  service for particle physics computing}},
  \href{https://doi.org/10.1007/s41781-019-0027-2}{\emph{Comput. Softw. Big
  Sci.} {\bfseries 3} (2019) 13}
  [\href{https://arxiv.org/abs/1904.08986}{{\ttfamily 1904.08986}}].

\bibitem{Krupa:2020bwg}
J.~Krupa et~al., \emph{{GPU coprocessors as a service for deep learning
  inference in high energy physics}},
  \href{https://doi.org/10.1088/2632-2153/abec21}{\emph{Mach. Learn. Sci.
  Tech.} {\bfseries 2} (2021) 035005}
  [\href{https://arxiv.org/abs/2007.10359}{{\ttfamily 2007.10359}}].

\bibitem{Rankin:2020usv}
D.S.~Rankin et~al., \emph{{FPGAs-as-a-Service Toolkit (FaaST)}},  in
  \emph{{2020 IEEE/ACM International Workshop on Heterogeneous High-performance
  Reconfigurable Computing (H2RC)}}, 10, 2020,
  \href{https://doi.org/10.1109/H2RC51942.2020.00010}{DOI}
  [\href{https://arxiv.org/abs/2010.08556}{{\ttfamily 2010.08556}}].

\bibitem{Wang:2020fjr}
M.~Wang, T.~Yang, M.~Acosta~Flechas, P.~Harris, B.~Hawks, B.~Holzman et~al.,
  \emph{{GPU-Accelerated Machine Learning Inference as a Service for Computing
  in Neutrino Experiments}},
  \href{https://doi.org/10.3389/fdata.2020.604083}{\emph{Front. Big Data}
  {\bfseries 3} (2021) 604083}
  [\href{https://arxiv.org/abs/2009.04509}{{\ttfamily 2009.04509}}].

\bibitem{Triton}
{Nvidia}, \emph{{Triton Inference Server}},  2022.

\bibitem{arrow}
{Apache}, ``{Apache Arrow}.'' \url{https://arrow.apache.org/docs/index.html},
  2022.

\bibitem{atlasStorage}
A.~Collaboration, \emph{{ATLAS Software and Computing HL-LHC Roadmap}},  Tech.
  Rep. \href{https://cds.cern.ch/record/2802918}{CERN-LHCC-2022-005,
  LHCC-G-182}, CERN, Geneva (Mar, 2022).

\bibitem{rootRNTuple}
J.~Lopez-Gomez and J.~Blomer, \emph{{RNTuple performance: Status and Outlook}},
   in \emph{{20th International Workshop on Advanced Computing and Analysis
  Techniques in Physics Research}: {AI Decoded - Towards Sustainable, Diverse,
  Performant and Effective Scientific Computing}}, 4, 2022
  [\href{https://arxiv.org/abs/2204.09043}{{\ttfamily 2204.09043}}].

\bibitem{hpcStorage}
A.~Bashyal, P.~Van~Gemmeren, S.~Sehrish, K.~Knoepfel, S.~Byna and Q.~Kang,
  \emph{Data storage for hep experiments in the era of high-performance
  computing},  Tech. Rep. \href{https://arxiv.org/abs/2203.07885}{} (2022),
  \href{https://doi.org/10.48550/ARXIV.2203.07885}{DOI}.

\bibitem{atlasNavigation}
{\scshape ATLAS} collaboration, \emph{{Next-Generation Navigational
  Infrastructure and the ATLAS Event Store}},
  \href{https://doi.org/10.1088/1742-6596/513/5/052036}{\emph{J. Phys. Conf.
  Ser.} {\bfseries 513} (2014) 052036}.

\bibitem{chakraborty:ccgrid22}
J.~Chakraborty, I.~Jimenez, S.A.~Rodriguez, A.~Uta, J.~LeFevre and C.~Maltzahn,
  \emph{Skyhook: Towards an arrow-native storage system},  in \emph{CCGrid22},
  (Taormina (Messina), Italy), May 16-19, 2022.

\bibitem{agc}
{IRIS-HEP}, ``{IRIS-HEP Analysis Grand Challenge}.''
  \url{https://iris-hep.org/grand-challenges.html}, 2022.

\bibitem{smith2020coffea}
N.~Smith, L.~Gray, M.~Cremonesi, B.~Jayatilaka, O.~Gutsche, A.~Hall et~al.,
  \emph{Coffea columnar object framework for effective analysis},  in \emph{EPJ
  Web of Conferences}, vol.~245, p.~06012, EDP Sciences, 2020.

\bibitem{benjamin2022analysis}
D.~Benjamin, K.~Bloom, B.~Bockelman, L.~Bryant, K.~Cranmer, R.~Gardner et~al.,
  \emph{Analysis facilities for hl-lhc}, {\emph{arXiv preprint
  arXiv:2203.08010} (2022) }.

\bibitem{flechas2022collaborative}
M.A.~Flechas, G.~Attebury, K.~Bloom, B.~Bockelman, L.~Gray, B.~Holzman et~al.,
  \emph{Collaborative computing support for analysis facilities exploiting
  software as infrastructure techniques}, {\emph{arXiv preprint
  arXiv:2203.10161} (2022) }.

\bibitem{lannon2022analysis}
K.~Lannon, P.~Brenner, M.~Hildreth, K.H.~Anampa, A.M.~Rodrigues, K.~Mohrman
  et~al., \emph{Analysis cyberinfrastructure: Challenges and opportunities},
  {\emph{arXiv preprint arXiv:2203.08811} (2022) }.

\bibitem{Feickert:2021sua}
M.~Feickert, L.~Heinrich, G.~Stark and B.~Galewsky, \emph{{Distributed
  statistical inference with pyhf enabled through funcX}},
  \href{https://doi.org/10.1051/epjconf/202125102070}{\emph{EPJ Web Conf.}
  {\bfseries 251} (2021) 02070}
  [\href{https://arxiv.org/abs/2103.02182}{{\ttfamily 2103.02182}}].

\bibitem{k8s}
{Kubernetes}, ``{Kubernetes}.'' \url{https://kubernetes.io/}, 2022.

\bibitem{bockelman2020wlcg}
B.~Bockelman, A.~Ceccanti, I.~Collier, L.~Cornwall, T.~Dack, J.~Guenther
  et~al., \emph{Wlcg authorisation from x. 509 to tokens},  in \emph{EPJ Web of
  Conferences}, vol.~245, p.~03001, EDP Sciences, 2020.

\bibitem{balcas2017cms}
J.~Balcas, B.~Bockelman, R.~Gardner, K.H.~Anampa, B.~Jayatilaka, F.A.~Khan
  et~al., \emph{Cms connect},  in \emph{Journal of Physics: Conference Series},
  vol.~898, p.~082032, IOP Publishing, 2017.

\bibitem{jh}
{JupyterHub}, ``{JupyterHub}.'' \url{https://jupyter.org/hub}, 2022.

\bibitem{HSFAF}
{HEP Software Foundation}, ``{HEP Software Foundation Analysis Facilities
  Forum}.''
  \url{https://hepsoftwarefoundation.org/activities/analysisfacilitiesforum.html},
  2022.

\bibitem{adamec2021coffea}
M.~Adamec, G.~Attebury, K.~Bloom, B.~Bockelman, C.~Lundstedt, O.~Shadura
  et~al., \emph{Coffea-casa: an analysis facility prototype},  in \emph{EPJ Web
  of Conferences}, vol.~251, p.~02061, EDP Sciences, 2021.

\bibitem{okd}
{OKD}, ``{OKD}.'' \url{https://www.okd.io/}, 2022.

\bibitem{ragan2018binder}
B.~Ragan-Kelley and C.~Willing, \emph{Binder 2.0-reproducible, interactive,
  sharable environments for science at scale},  in \emph{Proceedings of the
  17th Python in Science Conference (F. Akici, D. Lippa, D. Niederhut, and M.
  Pacer, eds.)}, pp.~113--120, 2018.

\bibitem{morris2016infrastructure}
K.~Morris, \emph{Infrastructure as code: managing servers in the cloud}, "
  O'Reilly Media, Inc." (2016).

\bibitem{beetz2021gitops}
F.~Beetz and S.~Harrer, \emph{Gitops: The evolution of devops?}, {\emph{IEEE
  Software} (2021) }.

\bibitem{OpenID}
{OpenID}, ``{OpenID}.'' \url{https://openid.net/what-is-openid/}, 2022.

\bibitem{GoogleID}
{Google}, ``{Google Identity}.'' \url{https://developers.google.com/identity/},
  2022.

\bibitem{ORCID}
{ORCID}, ``{ORCID}.'' \url{https://orcid.org}, 2022.

\bibitem{Spin}
{NERSC}, ``{Spin}.'' \url{https://www.nersc.gov/systems/spin/}, 2022.

\bibitem{NERSC}
{NERSC}, ``{NERSC}.'' \url{https://www.nersc.gov/}, 2022.

\bibitem{MyESnet}
\emph{Esnet volume history},  April, 2022.

\bibitem{ESnet-HEP-Requirements}
J.~Zurawski, B.~Brown, D.~Carder, E.~Colby, E.~Dart, K.~Miller et~al.,
  \emph{2020 high energy physics network requirements review final report},
  Report \href{https://escholarship.org/uc/item/78j3c9v4}{LBNL-2001398},
  Lawrence Berkeley National Laboratory (2021).

\bibitem{Shannon-Ciena}
B.~Lavall{\'{e}}e, \emph{Shannon’s limit, or opportunity?},  May, 2020.

\bibitem{Shannon-HuaWei}
J.~Yu, \emph{Approaching shannon’s limit: The way forward for optical
  transport},  May, 2020.

\bibitem{CoherentDetectLimit}
R.-J.~Essiambre, G.~Kramer, P.J.~Winzer, G.J.~Foschini and B.~Goebel,
  \emph{Capacity limits of optical fiber networks},
  \href{https://doi.org/10.1109/JLT.2009.2039464}{\emph{Journal of Lightwave
  Technology} {\bfseries 28} (2010) 662}.

\bibitem{LinuxTC}
``Introduction to linux traffic control.''

\bibitem{10.1145/3012426.3022184}
N.~Cardwell, Y.~Cheng, C.S.~Gunn, S.H.~Yeganeh and V.~Jacobson, \emph{Bbr:
  Congestion-based congestion control: Measuring bottleneck bandwidth and
  round-trip propagation time},
  \href{https://doi.org/10.1145/3012426.3022184}{\emph{Queue} {\bfseries 14}
  (2016) 20–53}.

\bibitem{bbrv2}
N.~Cardwell, Y.~Cheng, S.H.~Yeganeh, I.~Swett, V.~Vasiliev, P.~Jha et~al.,
  \emph{Bbrv2: A model-based congestion control},  in \emph{Presentation in
  ICCRG at IETF 104th meeting}, 2019.

\bibitem{M-21-07}
\emph{M-21-07: Completing the transition to internet protocol version 6
  (1pv6)},  November, 2020.

\bibitem{SegRouting}
\emph{{RFC} 8402 segment routing architecture},  July, 2018.

\bibitem{SegRtHead}
\emph{{RFC} 8754 ipv6 segment routing header (srh)},  March, 2020.

\bibitem{NOTED-FTS}
{Busse-Grawitz, Coralie}, {Martelli, Edoardo}, {Lassnig, Mario}, {Manzi,
  Andrea}, {Keeble, Oliver} and {Cass, Tony}, \emph{The noted software tool-set
  improves efficient network utilization for rucio data transfers via fts},
  \href{https://doi.org/10.1051/epjconf/202024507022}{\emph{EPJ Web Conf.}
  {\bfseries 245} (2020) 07022}.

\bibitem{Whitepaper-DataTrans-NetServices}
T.~Lehman, X.~Yang, C.~Guok, F.~Wuerthwein, I.~Sfiligoi, J.~Graham et~al.,
  \emph{Data transfer and network services management for domain science
  workflows},  2022.
\newblock 10.48550/ARXIV.2203.08280.

\bibitem{SuperFacilityAPI}
\emph{Superfacility api documentation},  2022.
\newblock https://docs.nersc.gov/services/sfapi/.

\bibitem{RARE}
\emph{Router for academia and research \& education},  2022.
\newblock https://wiki.geant.org/display/RARE/Home.

\bibitem{HECATE}
M.~Kiran, S.~Campbell and N.~Burgalio, \emph{Hecate: Towards self-driving
  networks in real-world},  November, 2021.
\newblock https://sc21.supercomputing.org/app/uploads/2021/11/SC21-NRE-001.pdf.

\bibitem{DOE-IRI-whitepaper}
B.~Brown, C.~Adams, K.~Antypas, B.~D, C.~S, E.Dart et~al., \emph{Toward a
  seamless integration of computing, experimental, and observational science
  facilities: A blueprint to accelerate discovery},  March, 2021.

\bibitem{RNTWG-firefly}
S.~McKee and M.~Babik, \emph{Packet and flow marking for global science
  domains},  September, 2021.
\newblock https://doi.org/10.5281/zenodo.6471024.

\bibitem{Kim2015InbandNT}
C.~Kim, A.~Sivaraman, N.P.K.~Katta, A.~Bas, A.A.~Dixit and L.J.~Wobker,
  \emph{In-band network telemetry via programmable dataplanes},  2015.

\bibitem{10.1145/3452411.3464443}
Z.~Liu, B.~Mah, Y.~Kumar, C.~Guok and R.~Cziva, \emph{Programmable per-packet
  network telemetry: From wire to kafka at scale},  in \emph{Proceedings of the
  2021 on Systems and Network Telemetry and Analytics}, SNTA '21, (New York,
  NY, USA), p.~33–36, Association for Computing Machinery, 2020,
  \href{https://doi.org/10.1145/3452411.3464443}{DOI}.

\bibitem{NetCacheDistScience}
A.~Sim, E.~Kissel and C.~Guok, \emph{Deploying in-network caches in support of
  distributed scientific data sharing},  Tech. Rep.
  \href{https://arxiv.org/abs/2203.06843}{} (2022),
  \href{https://doi.org/10.48550/ARXIV.2203.06843}{DOI}.

\bibitem{TransLayerNet}
Y.~Kumar, S.~Sheldon and D.~Carder, \emph{Transport layer networking},  Tech.
  Rep. \href{https://arxiv.org/abs/2204.02861}{} (2022),
  \href{https://doi.org/10.48550/ARXIV.2204.02861}{DOI}.

\bibitem{OSCARS}
C.~Guok, D.~Robertson, M.~Thompson, J.~Lee, B.~Tierney and W.~Johnston,
  \emph{Intra and interdomain circuit provisioning using the oscars reservation
  system},  in \emph{2006 3rd International Conference on Broadband
  Communications, Networks and Systems}, pp.~1--8, 2006,
  \href{https://doi.org/10.1109/BROADNETS.2006.4374316}{DOI}.

\bibitem{SENSE}
I.~Monga, C.Guok, J.~MacAuley, A.~Sim, H.~Newman, J.Balcas et~al.,
  \emph{Software-defined ntwork for end-to-end networked science at the
  exascale},  April, 2020,
  \href{https://doi.org/10.48550/arXiv.2004.05953}{DOI}.

\bibitem{RNTWG}
S.~McKee and M.~Babik, \emph{The research networking technical working group
  charter},  Charter
  \href{https://docs.google.com/document/d/1l4U5dpH556kCnoIHzyRpBl74IPc0gpgAG3VPUp98lo0/edit?usp=sharing}{}
  (2021), \href{https://doi.org/10.5281/zenodo.6470973}{DOI}.

\bibitem{RNTWG-packetmarking}
S.~McKee and M.~Babik, \emph{The research networking technical working group -
  packet marking sub group charter},  Charter
  \href{https://docs.google.com/document/d/1aAnsujpZnxn3oIUL9JZxcw0ZpoJNVXkHp-Yo5oj-B8U/edit?usp=sharing}{}
  (2021), \href{https://doi.org/10.5281/zenodo.6471050}{DOI}.

\bibitem{WLCG-DataLakes}
{Bird, Ian}, {Campana, Simone}, {Girone, Maria}, {Espinal, Xavier}, {McCance,
  Gavin} and {Schovancov\'a, Jaroslava}, \emph{Architecture and prototype of a
  wlcg data lake for hl-lhc},
  \href{https://doi.org/10.1051/epjconf/201921404024}{\emph{EPJ Web Conf.}
  {\bfseries 214} (2019) 04024}.

\bibitem{OSI-Model}
H.Z.~J.~Day, \emph{The osi reference model},  vol.~71, pp.~1334--1340,
  December, 1983, \href{https://doi.org/10.1109/PROC.1983.12775}{DOI}.

\bibitem{kandula2014calendaring}
S.~Kandula, I.~Menache, R.~Schwartz and S.R.~Babbula, \emph{Calendaring for
  wide area networks},  in \emph{Proceedings of the 2014 ACM conference on
  SIGCOMM}, pp.~515--526, 2014.

\bibitem{jalaparti2016dynamic}
V.~Jalaparti, I.~Bliznets, S.~Kandula, B.~Lucier and I.~Menache, \emph{Dynamic
  pricing and traffic engineering for timely inter-datacenter transfers},  in
  \emph{Proceedings of the 2016 ACM SIGCOMM Conference}, pp.~73--86, 2016.

\bibitem{mcclatchey2007data}
R.~McClatchey, A.~Anjum, H.~Stockinger, A.~Ali, I.~Willers and M.~Thomas,
  \emph{Data intensive and network aware (diana) grid scheduling},
  {\emph{Journal of Grid computing} {\bfseries 5} (2007) 43}.

\end{thebibliography}\endgroup
